\documentclass[prx,preprint,amsmath,superscriptaddress,nofootinbib,amssymb]{revtex4-2}
\usepackage{hyperref}
\usepackage{graphics}
\usepackage{amssymb}
\usepackage{upgreek}
\usepackage{graphicx}
\usepackage{times}
\usepackage{dcolumn}
\usepackage[usenames,dvipsnames]{color}
\usepackage{float}
\usepackage[utf8]{inputenc}

\usepackage[normalem]{ulem}
\usepackage{bm}
\usepackage{soul}
\usepackage{ragged2e}
\date{\today}

\begin{document}
\preprint{}

\title{Experimental demonstration of a magnetically induced warping transition in a topological insulator mediated by rare-earth surface dopants}

\author{Beatriz Mu\~{n}iz Cano}
\affiliation{Instituto Madrile\~no de Estudios Avanzados, IMDEA Nanociencia, Calle Faraday 9, 28049, Madrid, Spain}
\author{Yago Ferreiros}
\affiliation{Instituto Madrile\~no de Estudios Avanzados, IMDEA Nanociencia, Calle Faraday 9, 28049, Madrid, Spain}
\author{Pierre A. Pantale\'on}
\affiliation{Instituto Madrile\~no de Estudios Avanzados, IMDEA Nanociencia, Calle Faraday 9, 28049, Madrid, Spain}
\author{Ji Dai}
\affiliation{ALBA Synchrotron Light Source, 08290, Cerdanyola del Vall\`es, Barcelona, Spain}
\author{Massimo Tallarida}
\affiliation{ALBA Synchrotron Light Source, 08290, Cerdanyola del Vall\`es, Barcelona, Spain}
\author{Adriana I. Figueroa}
\affiliation{Departament de Física de la Matèria Condensada, Universitat de Barcelona, 08028 Barcelona, Spain.}
\affiliation{Catalan Institute of Nanoscience and
Nanotechnology (ICN2), CSIC and BIST, Campus UAB, 08193, Barcelona, Spain}
\author{Vera Marinova}
\affiliation{Institute of Optical Materials and Technologies, Bulgarian Academy of Sciences, Acad. G. Bontchev, Str. 109, 1113, Sofia, Bulgaria}
\author{Kevin García Díez}
\affiliation{ALBA Synchrotron Light Source, 08290, Cerdanyola del Vall\`es, Barcelona, Spain}
\affiliation{Catalan Institute of Nanoscience and
Nanotechnology (ICN2), CSIC and BIST, Campus UAB, 08193, Barcelona, Spain}
\author{Aitor Mugarza}
\affiliation{Catalan Institute of Nanoscience and
Nanotechnology (ICN2), CSIC and BIST, Campus UAB, 08193, Barcelona, Spain}
\affiliation{ICREA Instituci\'o Catalana de Recerca i Estudis Avan\c cats, Lluis Companys 23, 08010, Barcelona, Spain} 
\author{Sergio O. Valenzuela}
\affiliation{Catalan Institute of Nanoscience and
Nanotechnology (ICN2), CSIC and BIST, Campus UAB, 08193, Barcelona, Spain}
\affiliation{ICREA Instituci\'o Catalana de Recerca i Estudis Avan\c cats, Lluis Companys 23, 08010, Barcelona, Spain} 
\author{Rodolfo Miranda}
\affiliation{Instituto Madrile\~no de Estudios Avanzados, IMDEA Nanociencia, Calle Faraday 9, 28049, Madrid, Spain}
\affiliation{Departamento de F\'isica de la Materia Condensada; Instituto ``Nicol\'as Cabrera''
and Condensed Matter Physics Center (IFIMAC), Universidad Autónoma de Madrid (UAM), Campus de Cantoblanco, 28049, Madrid, Spain}
\author{Julio Camarero}
\affiliation{Instituto Madrile\~no de Estudios Avanzados, IMDEA Nanociencia, Calle Faraday 9, 28049, Madrid, Spain}
\affiliation{Departamento de F\'isica de la Materia Condensada; Instituto ``Nicol\'as Cabrera''
and Condensed Matter Physics Center (IFIMAC), Universidad Autónoma de Madrid (UAM), Campus de Cantoblanco, 28049, Madrid, Spain}
\author{Francisco Guinea}
\affiliation{Instituto Madrile\~no de Estudios Avanzados, IMDEA Nanociencia, Calle Faraday 9, 28049, Madrid, Spain}
\affiliation{Donostia International Physics Center, Paseo Manuel de Lardiz\'abal 4, 20018, San Sebasti\'an, Spain; and Ikerbasque, Basque Foundation for Science, 48009, Bilbao, Spain}
\author{Jose Angel Silva-Guill\'en}
\affiliation{Instituto Madrile\~no de Estudios Avanzados, IMDEA Nanociencia, Calle Faraday 9, 28049, Madrid, Spain}
\author{Miguel A. Valbuena}
\email{miguelangel.valbuena@imdea.org}
\affiliation{Instituto Madrile\~no de Estudios Avanzados, IMDEA Nanociencia, Calle Faraday 9, 28049, Madrid, Spain}

\begin{abstract}
Magnetic topological insulators (MTI) constitute a novel class of materials where the topologically protected band structure coexists with long-range ferromagnetic order, which can lead to the breaking of time-reversal symmetry (TRS), introducing a bandgap in the Dirac cone-shaped topological surface state (TSS). The gap opening in MITs has been predicted to be accompanied by a distortion in the TSS, evolving its warped shape from hexagonal to trigonal. In this work, we demonstrate such a transition by means of angle-resolved photoemission spectroscopy after the deposition of low concentrations of magnetic rare earths, namely Er and Dy, on the ternary three-dimensional prototypical topological insulator Bi$_2$Se$_2$Te. Signatures of the gap opening occurring as a consequence of the TRS breaking have also been observed, whose existence is supported by the observation of the aforementioned transition. Moreover, increasing the Er coverage results in a tunable p-type doping of the TSS. As a consequence, the Fermi level (E$_{\textrm{F}}$) of our Bi$_2$Se$_2$Te crystals can be gradually tuned towards the TSS Dirac point, and therefore to the magnetically induced bandgap; thus fulfilling two of the necessary prerequisites for the realization of the quantum anomalous Hall effect (QAHE) in this system. The experimental results are rationalized by a theoretical model where a magnetic Zeeman out-of-plane term is introduced in the hamiltonian governing the TSS band dispersion. Our results offer new strategies to control magnetic interactions with TSSs based on a simple approach and open up viable routes for the realization of the QAHE.
\end{abstract}

\maketitle

\normalsize
\newpage

\centering
\section{Introduction}
\justify

Topological insulators (TI) define a state of matter where the strong spin-orbit interaction (SOI) induces an exotic metallic topological surface state (TSS) with relativistic, Dirac-like, band dispersion and the spin locked to the momentum in an otherwise insulating material \cite{hasan2010colloquium,moore2010birth}. The interaction of the TSSs with magnetic impurities is especially attractive in these materials due to the emergence of novel quantum phenomena, all with relevant fundamental and technological implications in spintronics and quantum information processing \cite{liu2021magnetic}. The combination of topological properties and magnetic order can lead to new quantum states of matter as the quantum anomalous Hall effect (QAHE), characterized by purely spin-polarized dissipationless currents in the absence of an external magnetic field \cite{yu2010quantized,fei2020material}. 

When magnetism is introduced by impurity doping with magnetic elements, either via substitutional or surface dopants \cite{xu2012hedgehog,chen2010massive,sessi2016superparamagnetism,russmann2018towards}, by magnetic extension \cite{otrokov2017magnetic}, or by proximity coupling to magnetic layers \cite{kandala2013growth,hou2019magnetizing}, the TI becomes magnetic (MTI) and the time-reversal symmetry (TRS) can be broken, inducing the opening of a bandgap at the Dirac point (DP) of the TSS \cite{chen2010massive,xu2012hedgehog}. If this magnetically induced gap at the DP is tuned to the Fermi level (E$_\mathrm{F}$), the QAHE can be realized as it was first experimentally observed in Cr- and V- doped (Bi,Sb)$_2$Te$_3$ thin films \cite{chang2013experimental,chang2015high}. Recent discoveries based on the magnetic extension of TIs, as the first antiferromagnetic (AFM) TI, MnBi$_2$Te$_4$ \cite{otrokov2019prediction,rienks2019large}, have opened up new perspectives for the realization of these quantized topological effects in intrinsically magnetic stoichiometric compounds. However, these systems heavily depend on sophisticated and complex growth methods, resulting in variations in the density of structural defects, which can lead to drastic changes in the electronic structure of the TSSs (opening or not of an intrinsic magnetic gap at the DP) or affect its magnetic properties (formation or not of AFM order) \cite{garnica2022native}. Similar discrepancies concerning the magnetic order or full quantization of the observed QAHE on magnetic doped TIs have been found because of inhomogeneities of the spatial distribution of bulk magnetic dopants \cite{grauer2015coincidence,pan2020probing}. Thus, these quantum exotic phases only occur under certain extreme conditions of temperature, magnetic moment, anisotropy and exchange coupling with the surrounding electron bath \cite{liu2021magnetic,chang2013experimental,chang2015high,pan2020probing,eelbo2014strong}.

Surface doping offers an alternative to separate the two aforementioned critical processes, namely growth and doping, limiting the presence of dopants to the region where the TSSs are most sensitive to their effects and maximizing the magnetic anisotropy by the lower coordination symmetry \cite{gambardella2003giant}. Experimental attempts at doping the surface of TIs with magnetic species also indicate that achieving control on the magnetic ground state and anisotropy is challenging, mainly due to the existence of multiple adsorption sites, strong surface relaxations and significant electronic doping of bulk states from defects or inhomogeneities \cite{wray2011topological,scholz2012tolerance,valla2012photoemission,honolka2012plane}. Although gaplike features have been interpreted as magnetically induced, there are a number of other factors such as momentum and energy spatial fluctuations near the DP \cite{beidenkopf2011spatial} or surface chemical modifications \cite{chang2013thin,zhang2013topology,vobornik2014observation} that may contribute to the observation of a gap. Furthermore, the DP can be buried into the bulk band projections in such a way that the possible gap opening may be undetectable by techniques as angle resolved photoemission spectroscopy (ARPES). These ambiguities evidence the necessity to go beyond the present state-of-the-art in the development of these doping strategies \cite{cuxart2020molecular}.

An alternative route to this problem is the use of rare earths (RE) as surface dopants on TI substrates. Their larger size, as compared to other kind of dopants, can prevent the occurrence of substitutional sites at the surface, reducing the multiplicity of adsorption configurations \cite{abdalla2013topological} and enhancing the magnetic anisotropy.  In addition, REs large magnetic moments can maximize the magnetically induced gap that hosts the spin-polarized currents \cite{chen2010massive}. These large magnetic moments, originating from the unpaired 4f electrons \cite{jensen1991rare}, would also allow a lower doping concentration \cite{harrison2015study}, which could eventually lead to a more stable QAHE at higher temperatures \cite{liu2021magnetic}.  Besides their large magnetic moments and anisotropies, REs deposited on metals can be efficiently coupled via itinerant s and p electrons of the metal to achieve ferromagnetism with Curie temperatures as high as 80 K \cite{ormaza2016high,fernandez2020influence}. RE bulk substitutional doping with Eu \cite{fornari2020incorporation,tcakaev2020incipient}, Gd \cite{kim2015magnetic}, Dy \cite{harrison2015study} or Ho \cite{harrison2015massive,hesjedal2019rare} on Bi$_2$Te$_3$ thin films has been achieved. Interestingly, the TSSs were preserved despite the aforementioned REs large magnetic moments, nevertheless all films remained essentially paramagnetic or weak AFM. Regarding the bandgap opening, massive Dirac fermions have only been observed for a given concentration of Dy-doped Bi$_2$Te$_3$ thin films \cite{harrison2015massive}. Magnetic proximity effects have also been explored on EuS layers on TI thin films with neither significant induced magnetism within the TI nor an enhancement of the Eu magnetic moment at such interface \cite{figueroa2020absence}.

Very recently, it has been theoretically predicted that in MTIs, particularly in the newly discovered MnBi$_2$Te$_4$ family \cite{he2020mnbi2te4}, the magnetic order not only opens a gap at the DP but also lowers the symmetry of the Dirac cone warping from hexagonal to trigonal, together with an induction of a gap opening at the DP \cite{imai2021spintronic,naselli2022magnetic,tan2022momentum,wang2022three}. Notably, this trigonal warping at the magnetic ordered phase is sensitive to the direction of a net surface magnetic moment. Thus, this observation can provide an effective approach for the detection of magnetic ordering effects on the TSSs  in cases where the gap can not be unequivocally resolved and/or related to the TRS breaking by out-of-plane magnetic moments. However, an experimental observation of these phenomena is still lacking.

In this work, by means of ARPES, we report the effects on the TSS of the prototypical three-dimensional (3D) TI, Bi$_2$Se$_2$Te, after the deposition of RE impurities (Er and Dy). As expected, the pristine surface features hexagonal warped Dirac cones. As the RE atoms are deposited, even for small coverages, the TSSs exhibit the predicted transition from hexagonal to trigonal warping, providing the first experimental evidence of such magnetically induced warping modification. In addition, with the systematic Er (and Dy) doping, signatures of a bandgap opening at the DP are found, whereas the chemical potential is gradually modified resulting in a controllable p-type doping. This allows us to tune the DP energy closer to the E$_\mathrm{F}$, thus fulfilling the prerequisites for the realization of the QAHE. These observations have been rationalized in terms of the hamiltonian governing the TSS warping when a net magnetic moment is introduced, with an estimated exchange field coupling of $\approx$ 0.1eV.

 \centering
\section{Methods}
\justify
Ternary Bi$_2$Se$_2$Te single crystals \cite{wang2011ternary} were grown by a modified Bridgman method in a standard crystal growth system and characterized by X-ray powder diffraction (XRD) and Raman spectroscopy (see Supplemental Material (SM)~\cite{sm}), showing high crystal quality and crystalline long-range order. X-ray photoemission spectroscopy (XPS) and ARPES measurements were performed with an MBS hemispherical analyzer at the LOREA beamline \cite{garcia2022process,crisol2021alba} in Alba Synchrotron with linearly horizontal polarized light and photon energies of h$\upnu$ = 100 eV and 52 eV, respectively. Er (and Dy) were sublimated with an e-beam evaporator and the deposition rates and coverages were calibrated with a quartz crystal microbalance and then correlated with the attenuation of the Bi 5d core level on the XPS spectra (see SM\,\cite{sm}, Fig. S1). All XPS and ARPES measurements were performed at T = 15 K, well below the expected Curie temperature of Er and Dy clusters \cite{banister1954structure}, so a magnetic ordering of the RE surface dopants can be expected.

\centering
\section{Results and discussion}
\justify

\begin{figure}[H]
\centering
\includegraphics[width=\textwidth]{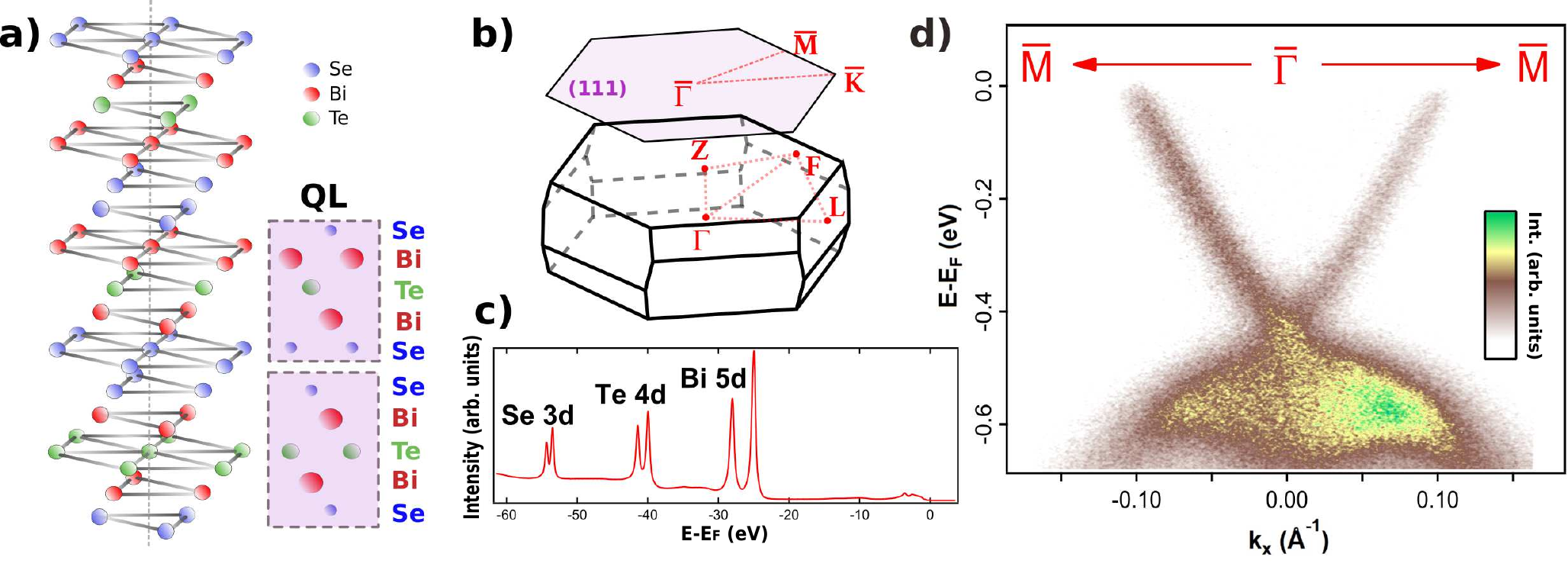}
	\caption{\textbf{Pristine Bi$_2$Se$_2$Te crystal structure and characterization.} (a) 3D quintuple layer crystal structure and (b) Brillouin zone (BZ) and projected surface BZ of Bi$_2$Se$_2$Te. (c) XPS spectrum (Se 3d, Te 4d, Bi 5d and valence band) of pristine Bi$_2$Se$_2$Te single crystal measured at a photon energy of h$\upnu$ = 100 eV and at T = 15 K. (d) ARPES bandmap of the pristine Bi$_2$Se$_2$Te TSS along the $\overline{\Gamma \text{M}}$ direction acquired at h$\upnu$ = 52 eV and at T = 15 K.}
	\label{Fig1_Bi2Se2Te}
\end{figure}

The Bi$_2$Se$_2$Te quintuple layer (QL) three dimensional (3D) crystal structure and the Brillouin 
zone (BZ) and its projected surface Brillouin zone (SBZ) are sketched in Figs.\,\ref{Fig1_Bi2Se2Te}(a)-(b), respectively. Bi$_2$Se$_2$Te bulk single crystals were mechanically exfoliated \textit{in-situ} with the sample held at 15 K in order to minimize the induced density of defects. The high sample quality is derived from the XPS spectrum (Fig.\,\ref{Fig1_Bi2Se2Te}(c)), showing highly intense and narrow peaks with no signs of oxidation, contamination or large disorder \cite{vobornik2014observation} for the Se 3d, Te 4d and Bi 5d core levels. The very well defined spin-orbit doublets as compared to isostructural TI films grown by molecular beam epitaxy prove the surface quality after \textit{in situ} exfoliation \cite{maass2014electronic,walsh2017interface,cuxart2020molecular}.

 The TSS ARPES bandmap of pristine Bi$_2$Se$_2$Te along the $\overline{\Gamma \text{M}}$ direction is shown in Fig. \,\ref{Fig1_Bi2Se2Te}(d). At this photon energy (h$\upnu$ = 52 eV) the bulk conduction band is expected to be above the E$_\mathrm{F}$. The DP is located at 415 meV below E$_\mathrm{F}$, outside the bulk projection bandgap, and the Fermi velocity is of v$_\mathrm{F}$ $ \sim (6.0 \pm 0.3) \cdot 10^{5}$ m/s, as estimated from $\text{1/}\hbar \text{ }(\partial \text{E}/ \partial \text{k})$ \cite{miyamoto2012topological}, both consistent with previous works \cite{miyamoto2012topological,bao2012weakARPES}.
 
\begin{figure}[H]
\centering	\includegraphics[width=
0.9\textwidth]{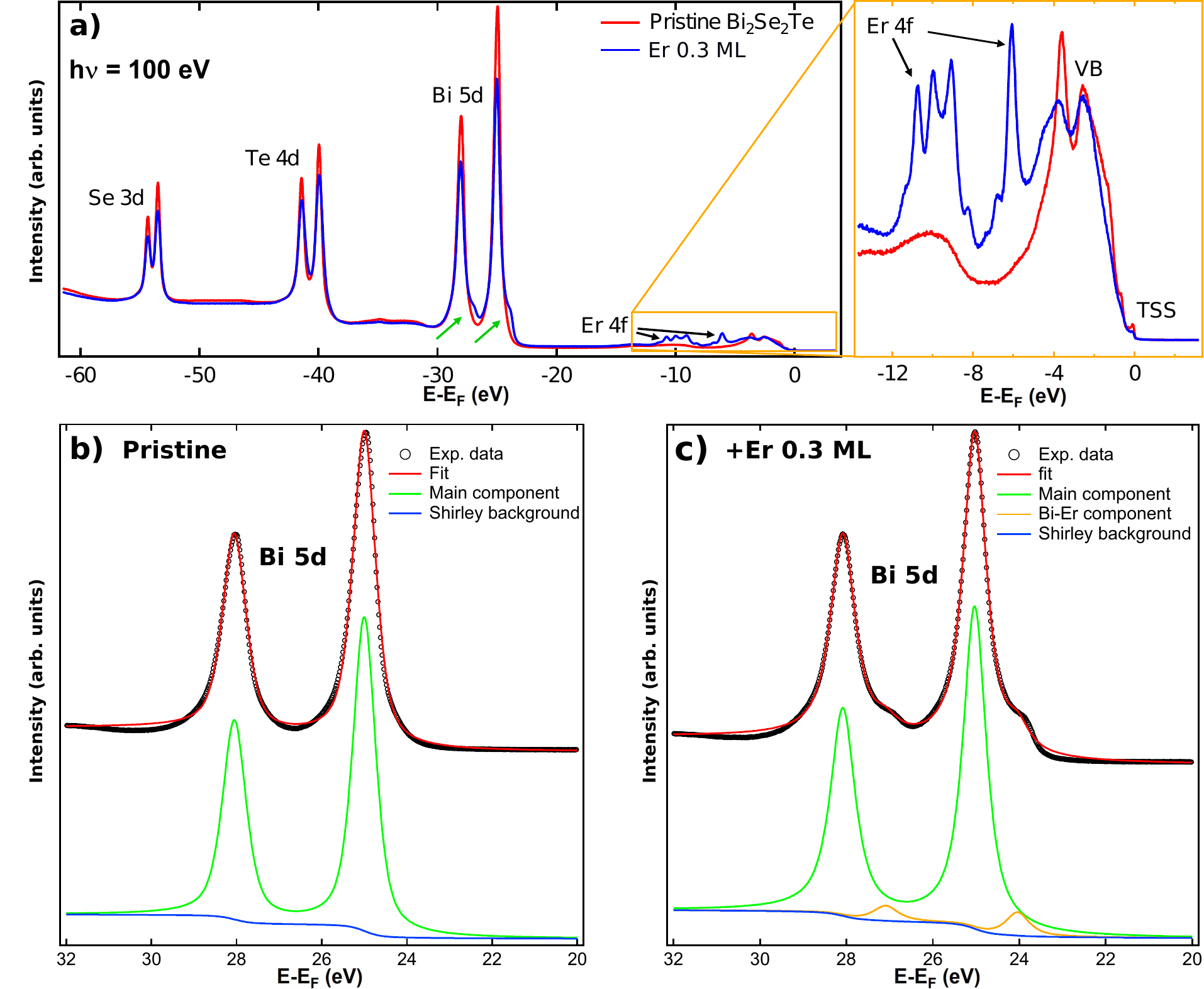}
\caption{\textbf{XPS characterization of Er-doped Bi$_2$Se$_2$Te acquired with a photon energy of h$\upnu$ = 100 eV and at T = 15K.} (a) Comparison of XPS spectra of pristine (red line) and 0.3 monolayer (ML) Er/Bi$_2$Se$_2$Te (blue line). A second component of the Bi 5d peak is developed after Er deposition (green arrows). Right: zoom-in on the Er 4f states and the Bi$_2$Se$_2$Te valence band energy range. Multipeak fit of XPS data for (b) the pristine sample and (c) the 0.3 ML Er-doped sample. A second component (orange) is related to Er-Bi interaction, as its area is proportional to the Er coverage (see SM \cite{sm}, Figs. S1 and S3).}
	\label{Fig2_XPS}
\end{figure}

Figure \ref{Fig2_XPS}(a) displays XPS spectra showing the main effects derived from the Er deposition on the core levels and the valence band (VB) states as compared to the pristine sample. The Se 3d, Te 4d and Bi 5d core levels are attenuated (see SM \cite{sm}, Fig. S3). No extra components are detected at higher binding energies (left side of peaks), consistent with the absence of surface oxidation or contamination. Additionally, no further disorder is introduced since no widening of the peak line-shapes is detected. This is also evidenced by the very sharp, well resolved multipeak structure of highly localized and, in principle, weakly interacting (non-dispersive) Er 4f states at E-E$_{\textrm{F}}$ = 5-12 eV (magnified in the inset in Fig.\,\ref{Fig2_XPS}(a)). The multipeak fit of the XPS spectra used for the estimation of the Er-coverage is shown in Figs.\,\ref{Fig2_XPS}(b)-(c). ARPES VB maps acquired at this energy region are included in the SM \cite{sm}, Fig. S2.

It is relevant to highlight the development of a second Bi 5d component at lower binding energies, indicated by green arrows in Fig.\,\ref{Fig2_XPS}(a) and in orange in Fig.\,\ref{Fig2_XPS}(c). The intensity of these features increases linearly with the Er coverage (see SM \cite{sm}, Figs. S1 and S3), which demonstrates its relationship with a superficial Bi-Er bond, and reveals a degree of interaction at the RE-TI interface. A charge transfer effect is also reflected in an energy shift as the Er content increases (see SM \cite{sm}, Fig. S1, green and orange dashed lines). A similar behavior has been observed for other metal-TI interfaces in the same Bi 5d state \cite{walsh2017interface}. Because of the large atomic sizes of Bi and RE dopants, Bi-Er bonds are most likely to be formed  than Se-Er or Te-Er ones, whose 3d and 4d states do not show extra components. The same Bi 5d second component also appears when Dy is deposited on Bi$_2$Se$_2$Te (see SM \cite{sm}, Fig. S5).

\begin{figure}[H]
\centering	\includegraphics[width=0.98\textwidth]{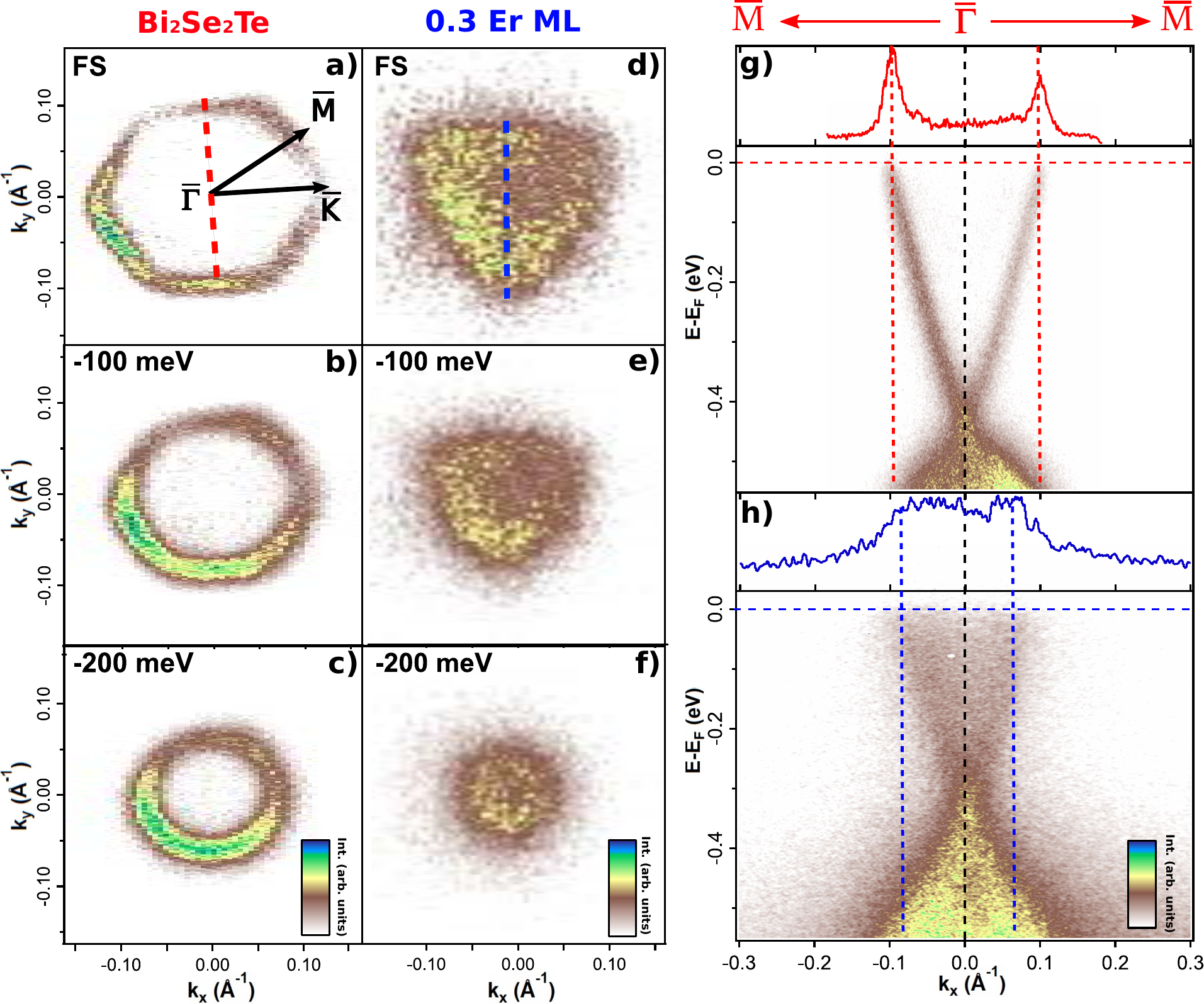}
	\caption{\textbf{Experimental observation by ARPES of the magnetically induced hexagonal to trigonal TSS warping transition upon Er deposition.} (a)-(c) Fermi surfaces and constante energy maps at 100 and 200 meV below the Fermi level (E$_\mathrm{F}$) for pristine Bi$_2$Se$_2$Te, showing the hexagonal warping of the topological surface state (TSS). (d)-(f) Same as in (a)-(c) for 0.3 monolayer (ML) Er/Bi$_2$Se$_2$Te, showing the trigonal-warping of the TSS. TSS bandmaps along the $\overline{\Gamma \text{M}}$ direction (red and blue dashed lines in (a) and (d)) for (g) pristine and (h) 0.3 ML Er/Bi$_2$Se$_2$Te, acquired with a photon energy h$\upnu$ = 52 eV and at T = 15  K. Momentum distribution curves (MDCs) extracted at the E$_\mathrm{F}$ are shown on top of each bandmap. The induction of a trigonal warping in the TSS is also evidenced by the Fermi wavevector k$_\mathrm{F}$ and Fermi velocity v$_\mathrm{F}$ asymmetry around the $\overline{\Gamma}$ point for the Er-doped system in (h).}
	\label{Fig3_FS}
\end{figure}

The Fermi surface (FS) and constant energy (CE) maps for pristine and 0.3 monolayer (ML) Er/Bi$_2$Se$_2$Te are shown in Fig.\,\ref{Fig3_FS}. Pristine Bi$_2$Se$_2$Te TSS exhibits the hexagonal warped FS and the circular-like shaped lower energy CE maps expected for rombohedral 3D TIs (Figs.\,\ref{Fig3_FS}(a)-(c)). The warping strength is consistent with the literature, being lower than in most warped TSSs, as in Bi$_2$Te$_3$ \cite{fu2009hexagonal,chen2009experimental}, and higher than in isostructural Bi$_2$Te$_2$Se \cite{arakane2012tunable} or Bi$_2$Se$_3$ \cite{xia2009observation}, whose warping effects are smoother or even negligible, respectively. When Er is deposited, the symmetry of the FS and CE maps changes drastically. Upon 0.3 Er ML, the TSS warping symmetry evolves from hexagonal to trigonal (Figs.\,\ref{Fig3_FS}(d)-(f)), which becomes clearer away from the DP and closer to E$_\mathrm{F}$. Similar modifications in the Dirac cone warping, from hexagonal to trigonal, have been predicted below the magnetic ordering temperature in the family of MnBi$_{\text{2n}}$Te$_{\text{3n+1}}$ MTIs \cite{naselli2022magnetic}. 

\begin{figure}[H]
\includegraphics[width=\textwidth]{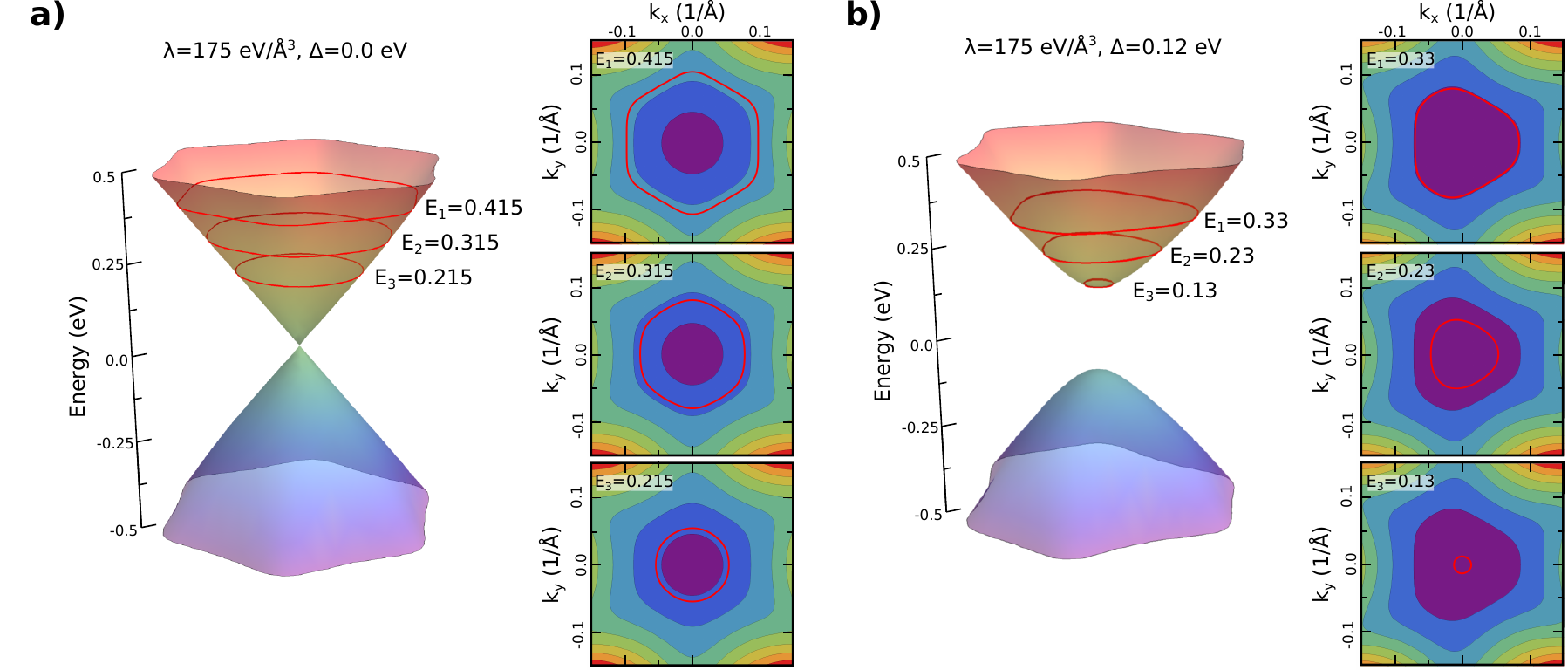}
\caption{\textbf{Theoretical modelling of the TSS band structure and the warping transition.} Band structure of the surface state of a topological insulator and the evolution of the constant energy (CE) maps (red lines) as a function of the Fermi energy for the (a) pristine system, with v$_\mathrm{F}=6.0 \cdot 10^{5}$ and $\uplambda =$ 175~eV/\AA,~and (b) the system with magnetic impurities, with  $\Delta=0.12$~eV. The transition from a six-fold to a trigonal warping, as well as the induction of a gap, is observed. The small panels display CE maps at three different energies relative to the Dirac point, corresponding to the energies of the experimental ones}
\label{Figtw}
\end{figure}

To give some insight into the warping transition of the TSS, the description of an undoped TI using the model developed by Fu~\cite{fu2009hexagonal} can be applied, in which the low energy dispersion of the system is described by the following Hamiltonian:
\begin{equation}
    \mathrm{H = v(k_x\upsigma_y-k_y\upsigma_x)+\frac{\uplambda}{2}(k_+^3+k_-^3)\upsigma_z},
    \label{eq:hamiltonianTI1}
\end{equation}
with $\mathrm{k_\pm=k_x\pm ik_y}$. 
Note that the last term is only invariant under threefold rotation and is responsible for the hexagonal warping~\cite{fu2009hexagonal,imai2021spintronic,naselli2022magnetic}.
The 3D band dispersion of the system is shown in Fig.~\ref{Figtw}(a), where the previously calculated v$_\mathrm{F}=6.0 \cdot 10^{5}$ m/s has been used to match the experimental value.
The warping parameter $\uplambda$ has been set to 175~eV/\AA~by fitting the theoretically calculated CE map at 0.415 eV (see right-hand side of Fig.~\ref{Figtw}(a)) to the measured CE map (Fig.\ref{Fig3_FS}(a)).
This value confirms a lower warping strength compared to other TIs such as Bi$_2$Te$_3$~\cite{fu2009hexagonal}.
The given values for the energies (measured with respect to the DP) in the theoretical CE maps in Fig.~\ref{Figtw}(a) have been chosen to match the experimental values in Figs.~\ref{Fig3_FS}(a)-(c).
As expected, the hexagonal warping can be clearly distinguished at energies far from the DP.
As the energy approaches the DP, the CE maps become more rounded, as seen in the experiment (Figs.~\ref{Fig3_FS}(a)-(c)).

To introduce the effect of the RE dopants in the former Hamiltonian (Eq.\,\ref{eq:hamiltonianTI1}), a magnetic moment can be coupled to the TI TSS via an exchange interaction~\cite{naselli2022magnetic} in what constitutes a Zeeman-like term. The Hamiltonian takes the final form (see SM \cite{sm}):

\begin{equation}
   \mathrm{H= v_F(k_x\upsigma_y-k_y\upsigma_x)+(\uplambda k^3\cos3\uptheta-\Delta)\upsigma_z},
\end{equation}
where $\Delta$ stands for the exchange coupling and $\uptheta$ for the azimuthal angle of momentum.
$\Delta$ is a fitting parameter of our model.
By fixing v$_{\text{F}}$ and $\uplambda$, a value of $\Delta=0.12$~eV is obtained by comparing the theoretical CE maps (right-hand side of Fig.~\ref{Figtw}(b)) to the experimental ones Figs.~\ref{Fig3_FS}(d)-(f). Note that the E$_{\text{F}}$ (with respect to the DP) is modified by the introduction of the RE dopants (see Figs.~\ref{Fig3_FS}(g)-(h)), so the values of the energies for the theoretical CE maps have been accordingly changed.
The energy dispersion of the system with the above set of parameters is shown on the left-hand side of Fig.~\ref{Figtw}(b), with the TSS band presenting a gap of 240 meV. This is a consequence of the non-vanishing net magnetization induced by the magnetic dopant.
Furthermore, the TRS is broken and the surface of the TI becomes a quantum anomalous Hall (or Chern) insulator \cite{qi2006topological,hasan2010colloquium}. As can be seen on CE maps (right-hand side of Fig.~\ref{Figtw}(b)), the breakdown of TRS results in the transition from the six-fold to a three-fold rotational symmetry (see SM \cite{sm}), so that the warped FS evolves from hexagonal (pristine system) to trigonal (doped system), just as in the experimental measurements (Figs.~\ref{Fig3_FS}(d)-(f)). Interestingly, the obtained value of $\Delta$ is up to one order of magnitude larger than most previous expectations~\cite{NN10,yokoyama2010theoretical,Y11,ferreiros2015dirac}.
Although higher values have also been predicted or expected~\cite{liu2009magnetic,FC14}, up to our knowledge, this is the first time that the exchange coupling has been estimated by direct comparison to experimental results.
Such a relatively high value of the exchange coupling $\Delta$ makes Bi$_2$Se$_2$Te doped with RE a good candidate system for the realization of the QAHE.

Experimentally, the change in the warping symmetry is also evidenced by the induced inversion asymmetry in the TSS band dispersion in Figs.\,\ref{Fig3_FS}(g) and (h), measured along the $\overline{\Gamma \text{M}}$ direction (red and blue dashed lines in Figs.\, \ref{Fig3_FS}(a) and (d)), and in the the momentum distribution curves (MDCs) close to the E$_\mathrm{F}$, depicted above each TSS bandmap. The Fermi wavevectors k$_\mathrm{F}$ for 0.3 ML Er/Bi$_2$Se$_2$Te (blue MDC, Fig.\,\ref{Fig3_FS}(h)) are asymmetric with respect to the $\overline{\Gamma}$ point, as compared to pristine Bi$_2$Se$_2$Te (red MDC, Fig.\,\ref{Fig3_FS}(g)), as a consequence of the different v$_{\text{F}}$ of the two branches of the TSSs. A similar asymmetric band dispersion of the TSS has been predicted for the AFM state of the MnBi$_2$Te$_4$(0001) surface following the same hexagonal-to-trigonal TSS transition \cite{tan2022momentum}.



\begin{figure}[H]
\centering
    \includegraphics[width=0.5\textwidth]{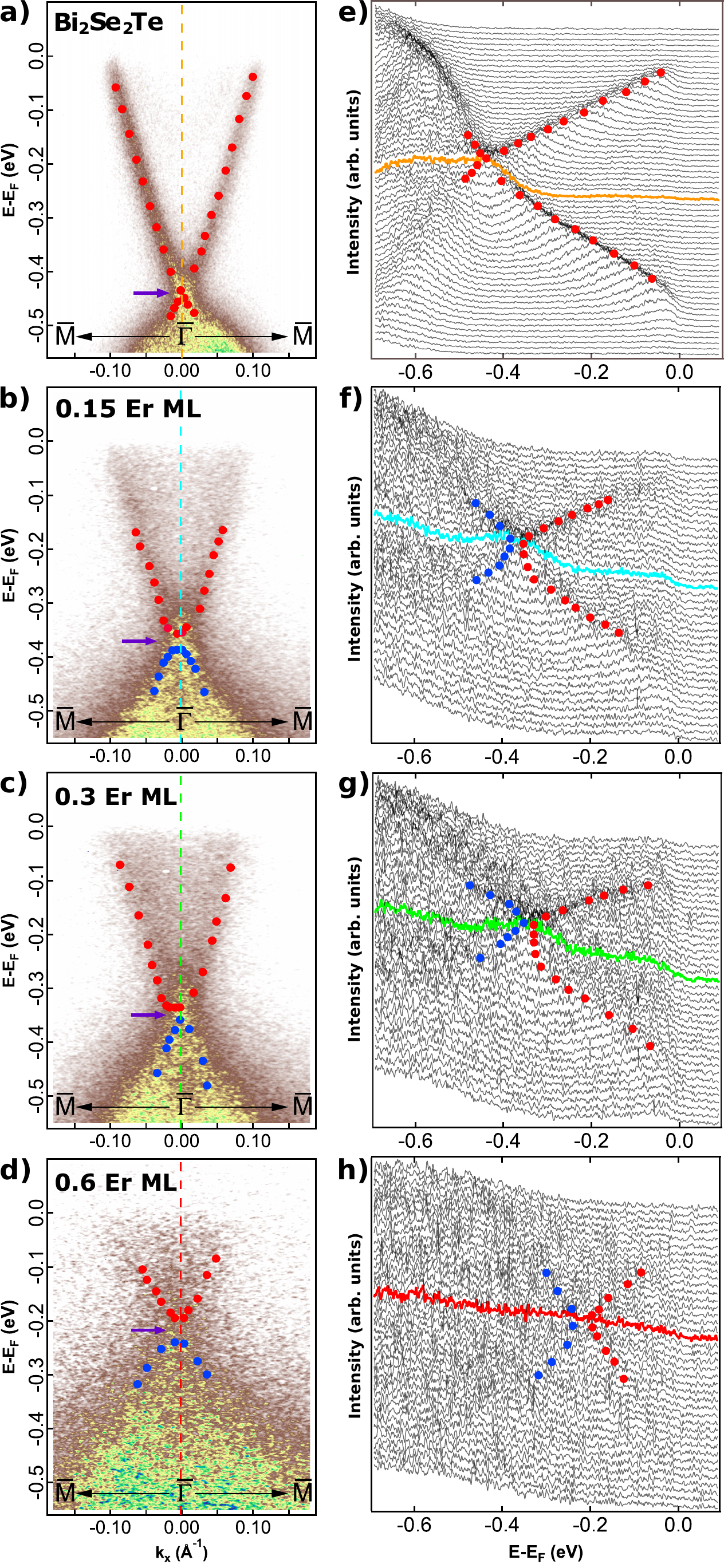}
	\caption{\textbf{Topological surface state (TSS) band structure evolution upon Er doping showing signatures of the magnetically induced bandgap at the Dirac point.} (a)-(d) TSS bandmaps for pristine Bi$_2$Se$_2$Te and 0.15, 0.3 and 0.6 Er monolayer (ML)/Bi$_2$Se$_2$Te, respectively, acquired with a photon energy of h$\upnu$ = 52 eV and at T = 15 K. (e)-(f) EDCs extracted from the TSS bandmaps in (a)-(d). EDCs at the $\overline{\Gamma}$ point are highlighted following the color correspondence. TSS dispersions are obtained by fitting the EDCs and are indicated by blue and red dots for upper and bottom bands with respect to the DP, respectively. Purple arrows indicate the DP energy, which is upshifted as the Er coverage is increased.}
    \label{Fig4_ARPES_TSS}
\end{figure}

The evolution of the TSS as a function of the Er coverage is shown in Fig.\,\ref{Fig4_ARPES_TSS}. TSS ARPES bandmaps along the $\overline{\Gamma \text{M}}$ direction for pristine, 0.15, 0.3 and 0.6 ML Er/Bi$_2$Se$_2$Te are shown in Figs.\,\ref{Fig4_ARPES_TSS}(a)-(d), respectively. The corresponding energy distribution curves (EDCs) as a function of the Er coverage are presented in Figs.\,\ref{Fig4_ARPES_TSS}(e)-(h), where the EDCs acquired precisely at the $\overline{\Gamma}$ point and thus crossing the DP (along vertical dashed lines in Figs.\,\ref{Fig4_ARPES_TSS}(a)-(d)) are highlighted following the color correspondence. Here, orange, cyan, green, and red lines stand for pristine, 0.15, 0.3, and 0.6 ML Er/Bi$_2$Se$_2$Te, respectively. The band dispersions have been obtained from the EDCs fitting, whose results are indicated by red and blue dots in Fig.\,\ref{Fig4_ARPES_TSS}. The dispersion obtained from the pristine sample (Figs.\,\ref{Fig4_ARPES_TSS}(a) and (e)) displays linear branches which clearly cross at the DP. Red and blue dots for the doped samples (Figs.\,\ref{Fig4_ARPES_TSS}(b)-(h)), indicate the upper and bottom band dispersions relative to the DP.

As previously described, the transition from hexagonal to trigonal of the TSS symmetry when an out-of-plane magnetic moment is introduced (and predicted in AFM/TIs systems \cite{naselli2022magnetic}) is expected to be accompanied by a bandgap opening at the DP because of the TRS breaking. A more detailed and comprehensive picture on the TSS band dispersion around the DP can be gathered from the EDCs fits (Figs.\,\ref{Fig4_ARPES_TSS}(e)-(h)); particularly, in those obtained at the $\overline{\Gamma}$ point, extracted along vertical dashed lines in Figs.\,\ref{Fig4_ARPES_TSS}(a)-(d), and highlighted in the same colors as in Fig.\,\ref{Fig4_ARPES_TSS}(e)-(h). For the pristine sample (Fig.\,\ref{Fig4_ARPES_TSS}(e)) the DP is clearly defined by the intersection of the two linear dispersive branches (orange) resulting in a single, punctual, sharp maximum of intensity at $\sim$ 415 meV. After the Er deposition, a broadening of the intensity is clearly evidenced around the DP, which would be an indication of a possible bandgap opening at the DP. This can be more clearly seen by following the band dispersion from the EDCs (red and blue dots in Figs.\,\ref{Fig4_ARPES_TSS}(f)-(g)). The characteristic linear dispersion of relativistic massless Dirac fermions of the pristine sample seems to be renormalized to a parabolic-like dispersion, showing a plateau in the EDC intensity in the vicinity of the $\overline{\Gamma}$ point. This would be an indication of the bandgap between the top of the bulk valence band and the bottom of the surface state. Similar XPS and ARPES results have been found after 0.3 Dy ML deposition and are summarized in the SM  \cite{sm}, Fig.S6.

As discussed at the beginning of the manuscript, similar bandgap openings have already been experimentally observed and attributed to TRS by magnetic impurities for numerous transition metal magnetic impurities \cite{chen2010massive} or Dy substitutional dopants \cite{harrison2015massive}. Nevertheless, the magnetically induced warping transition also resulting from the TRS breaking has not been reported so far.


\centering
\section{Conclusions}
\justify

In this work, by doping the surface of a TI with RE atoms, the magnetically induced evolution of the warping of the TSS from hexagonal to trigonal has been demonstrated by means of ARPES experiments. This shape transition has been predicted to be accompanied by a bandgap opening at the DP and, indeed, signatures of such effect have also been observed in the present study. Additionally, the DP has been proven to be tunable towards the E$_{\text{F}}$ through charge transfer between the RE adatoms and the substrate. The TSS band dispersion of the doped system is reproduced by including a Zeeman-like term in the low energy Hamiltonian describing the system. This modification entails two important consequences, being the first of them the just above-described breaking of the hexagonal symmetry and, the second one, the opening of a gap at the DP of the TSS. Both phenomena have been observed in the ARPES measurements. Moreover, we find that, in order to theoretically describe the experimental results, a large exchange coupling is needed, of the order of 0.1 eV.  Thus, the large exchange coupling and the DP tunability make the controlled doping of TIs with REs an excellent approach to realize the QAHE at higher temperatures.

\bigskip
\centering
\section*{Acknowledgments}   
\justify

IMDEA Nanociencia and IFIMAC acknowledge financial support from the Spanish Ministry of Science and Innovation through, respectively, ‘Severo Ochoa’ (Grant CEX2020-001039-S) and ‘María de Maeztu’ (Grant CEX2018-000805-M) Programmes for Centres of Excellence in R\&D. This project has received funding from the Community of Madrid (CM) through project P2018/NMT-4321 (NANOMAGCOST), from Spanish Ministry of Economic Affairs and Digital Transformation (MINECO) through projects SPGC2018-098613-B-C21 (SpOrQuMat),  EQC2019-006304-P (Equipamiento científico) and PID2020-116181RB-C31 (SOnanoBRAIN), and from Spanish Ministry of Science and Innovation (MICINN) and the Spanish Research Agency (AEI/10.13039/501100011033) through grants PID2021-123776NB-C21 (CONPHASE$^{\mathrm{TM}}$), PID2019-107338RB-C65 and PID2019-111773RB-I00. ICN2 was funded by the CERCA Programme/Generalitat de Catalunya and supported by the Spanish Ministry of Economy and Competitiveness, MINECO under Contract No. SEV-2017-0706. A.I.F., V.M. and S.O.V. acknowledge support from European Union's Horizon 2020 FET-PROACTIVE project TOCHA under Grant Agreement No. 824140. LOREA was co-funded by the European Regional Development Fund (ERDF) within the Framework of the Smart Growth Operative Programme 2014-2020. B.M.C. acknowledges support from CM (PEJD-2019-PRE/IND-17048). F.G. and P.A.P. acknowledge support from the European Commission, within the Graphene Flagship, Core 3 (Grant No. 881603) and from CM (Grant NMAT2D). Y.F. acknowledges financial support through the Programa de Atraccion de Talento de la Comunidad de Madrid, Spain, Grant No. 2018-T2/IND-11088. A.I.F. is a Serra Húnter fellow.

\bibliography{REs_TIs_warping.bib}

\begin{thebibliography}{66}%
\makeatletter
\providecommand \@ifxundefined [1]{%
 \@ifx{#1\undefined}
}%
\providecommand \@ifnum [1]{%
 \ifnum #1\expandafter \@firstoftwo
 \else \expandafter \@secondoftwo
 \fi
}%
\providecommand \@ifx [1]{%
 \ifx #1\expandafter \@firstoftwo
 \else \expandafter \@secondoftwo
 \fi
}%
\providecommand \natexlab [1]{#1}%
\providecommand \enquote  [1]{``#1''}%
\providecommand \bibnamefont  [1]{#1}%
\providecommand \bibfnamefont [1]{#1}%
\providecommand \citenamefont [1]{#1}%
\providecommand \href@noop [0]{\@secondoftwo}%
\providecommand \href [0]{\begingroup \@sanitize@url \@href}%
\providecommand \@href[1]{\@@startlink{#1}\@@href}%
\providecommand \@@href[1]{\endgroup#1\@@endlink}%
\providecommand \@sanitize@url [0]{\catcode `\\12\catcode `\$12\catcode
  `\&12\catcode `\#12\catcode `\^12\catcode `\_12\catcode `\%12\relax}%
\providecommand \@@startlink[1]{}%
\providecommand \@@endlink[0]{}%
\providecommand \url  [0]{\begingroup\@sanitize@url \@url }%
\providecommand \@url [1]{\endgroup\@href {#1}{\urlprefix }}%
\providecommand \urlprefix  [0]{URL }%
\providecommand \Eprint [0]{\href }%
\providecommand \doibase [0]{https://doi.org/}%
\providecommand \selectlanguage [0]{\@gobble}%
\providecommand \bibinfo  [0]{\@secondoftwo}%
\providecommand \bibfield  [0]{\@secondoftwo}%
\providecommand \translation [1]{[#1]}%
\providecommand \BibitemOpen [0]{}%
\providecommand \bibitemStop [0]{}%
\providecommand \bibitemNoStop [0]{.\EOS\space}%
\providecommand \EOS [0]{\spacefactor3000\relax}%
\providecommand \BibitemShut  [1]{\csname bibitem#1\endcsname}%
\let\auto@bib@innerbib\@empty
\bibitem [{\citenamefont {Hasan}\ and\ \citenamefont
  {Kane}(2010)}]{hasan2010colloquium}%
  \BibitemOpen
  \bibfield  {author} {\bibinfo {author} {\bibfnamefont {M.~Z.}\ \bibnamefont
  {Hasan}}\ and\ \bibinfo {author} {\bibfnamefont {C.~L.}\ \bibnamefont
  {Kane}},\ }\bibfield  {title} {\bibinfo {title} {\textit{Colloquium:
  Topological insulators}},\ }\href
  {https://doi.org/10.1103/RevModPhys.82.3045} {\bibfield  {journal} {\bibinfo
  {journal} {Reviews of Modern Physics}\ }\textbf {\bibinfo {volume} {82}},\
  \bibinfo {pages} {3045} (\bibinfo {year} {2010})}\BibitemShut {NoStop}%
\bibitem [{\citenamefont {Moore}(2010)}]{moore2010birth}%
  \BibitemOpen
  \bibfield  {author} {\bibinfo {author} {\bibfnamefont {J.~E.}\ \bibnamefont
  {Moore}},\ }\bibfield  {title} {\bibinfo {title} {\textit{The birth of
  topological insulators}},\ }\href {https://doi.org/10.1038/nature08916}
  {\bibfield  {journal} {\bibinfo  {journal} {Nature}\ }\textbf {\bibinfo
  {volume} {464}},\ \bibinfo {pages} {194} (\bibinfo {year}
  {2010})}\BibitemShut {NoStop}%
\bibitem [{\citenamefont {Liu}\ and\ \citenamefont
  {Hesjedal}(2021)}]{liu2021magnetic}%
  \BibitemOpen
  \bibfield  {author} {\bibinfo {author} {\bibfnamefont {J.}~\bibnamefont
  {Liu}}\ and\ \bibinfo {author} {\bibfnamefont {T.}~\bibnamefont {Hesjedal}},\
  }\bibfield  {title} {\bibinfo {title} {\textit{Magnetic topological insulator
  heterostructures: A review}},\ }\href
  {https://doi.org/10.1002/adma.202102427} {\bibfield  {journal} {\bibinfo
  {journal} {Advanced Materials}\ ,\ \bibinfo {pages} {2102427}} (\bibinfo
  {year} {2021})}\BibitemShut {NoStop}%
\bibitem [{\citenamefont {Yu}\ \emph {et~al.}(2010)\citenamefont {Yu},
  \citenamefont {Zhang}, \citenamefont {Zhang}, \citenamefont {Zhang},
  \citenamefont {Dai},\ and\ \citenamefont {Fang}}]{yu2010quantized}%
  \BibitemOpen
  \bibfield  {author} {\bibinfo {author} {\bibfnamefont {R.}~\bibnamefont
  {Yu}}, \bibinfo {author} {\bibfnamefont {W.}~\bibnamefont {Zhang}}, \bibinfo
  {author} {\bibfnamefont {H.-J.}\ \bibnamefont {Zhang}}, \bibinfo {author}
  {\bibfnamefont {S.-C.}\ \bibnamefont {Zhang}}, \bibinfo {author}
  {\bibfnamefont {X.}~\bibnamefont {Dai}},\ and\ \bibinfo {author}
  {\bibfnamefont {Z.}~\bibnamefont {Fang}},\ }\bibfield  {title} {\bibinfo
  {title} {\textit{Quantized anomalous Hall effect in magnetic topological
  insulators}},\ }\href {https://doi.org/10.1126/science.1187485} {\bibfield
  {journal} {\bibinfo  {journal} {Science}\ }\textbf {\bibinfo {volume}
  {329}},\ \bibinfo {pages} {61} (\bibinfo {year} {2010})}\BibitemShut
  {NoStop}%
\bibitem [{\citenamefont {Fei}\ \emph {et~al.}(2020)\citenamefont {Fei},
  \citenamefont {Zhang}, \citenamefont {Zhang}, \citenamefont {Shah},
  \citenamefont {Song}, \citenamefont {Wang},\ and\ \citenamefont
  {Wang}}]{fei2020material}%
  \BibitemOpen
  \bibfield  {author} {\bibinfo {author} {\bibfnamefont {F.}~\bibnamefont
  {Fei}}, \bibinfo {author} {\bibfnamefont {S.}~\bibnamefont {Zhang}}, \bibinfo
  {author} {\bibfnamefont {M.}~\bibnamefont {Zhang}}, \bibinfo {author}
  {\bibfnamefont {S.~A.}\ \bibnamefont {Shah}}, \bibinfo {author}
  {\bibfnamefont {F.}~\bibnamefont {Song}}, \bibinfo {author} {\bibfnamefont
  {X.}~\bibnamefont {Wang}},\ and\ \bibinfo {author} {\bibfnamefont
  {B.}~\bibnamefont {Wang}},\ }\bibfield  {title} {\bibinfo {title}
  {\textit{The material efforts for quantized Hall devices based on topological
  insulators}},\ }\href {https://doi.org/10.1002/adma.201904593} {\bibfield
  {journal} {\bibinfo  {journal} {Advanced Materials}\ }\textbf {\bibinfo
  {volume} {32}},\ \bibinfo {pages} {1904593} (\bibinfo {year}
  {2020})}\BibitemShut {NoStop}%
\bibitem [{\citenamefont {Xu}\ \emph {et~al.}(2012)\citenamefont {Xu},
  \citenamefont {Neupane}, \citenamefont {Liu}, \citenamefont {Zhang},
  \citenamefont {Richardella}, \citenamefont {Andrew~Wray}, \citenamefont
  {Alidoust}, \citenamefont {Leandersson}, \citenamefont {Balasubramanian},
  \citenamefont {S{\'a}nchez-Barriga} \emph {et~al.}}]{xu2012hedgehog}%
  \BibitemOpen
  \bibfield  {author} {\bibinfo {author} {\bibfnamefont {S.-Y.}\ \bibnamefont
  {Xu}}, \bibinfo {author} {\bibfnamefont {M.}~\bibnamefont {Neupane}},
  \bibinfo {author} {\bibfnamefont {C.}~\bibnamefont {Liu}}, \bibinfo {author}
  {\bibfnamefont {D.}~\bibnamefont {Zhang}}, \bibinfo {author} {\bibfnamefont
  {A.}~\bibnamefont {Richardella}}, \bibinfo {author} {\bibfnamefont
  {L.}~\bibnamefont {Andrew~Wray}}, \bibinfo {author} {\bibfnamefont
  {N.}~\bibnamefont {Alidoust}}, \bibinfo {author} {\bibfnamefont
  {M.}~\bibnamefont {Leandersson}}, \bibinfo {author} {\bibfnamefont
  {T.}~\bibnamefont {Balasubramanian}}, \bibinfo {author} {\bibfnamefont
  {J.}~\bibnamefont {S{\'a}nchez-Barriga}}, \emph {et~al.},\ }\bibfield
  {title} {\bibinfo {title} {\textit{Hedgehog spin texture and Berry’s phase
  tuning in a magnetic topological insulator}},\ }\href
  {https://doi.org/10.1038/nphys2351} {\bibfield  {journal} {\bibinfo
  {journal} {Nature Physics}\ }\textbf {\bibinfo {volume} {8}},\ \bibinfo
  {pages} {616} (\bibinfo {year} {2012})}\BibitemShut {NoStop}%
\bibitem [{\citenamefont {Chen}\ \emph {et~al.}(2010)\citenamefont {Chen},
  \citenamefont {Chu}, \citenamefont {Analytis}, \citenamefont {Liu},
  \citenamefont {Igarashi}, \citenamefont {Kuo}, \citenamefont {Qi},
  \citenamefont {Mo}, \citenamefont {Moore}, \citenamefont {Lu} \emph
  {et~al.}}]{chen2010massive}%
  \BibitemOpen
  \bibfield  {author} {\bibinfo {author} {\bibfnamefont {Y.}~\bibnamefont
  {Chen}}, \bibinfo {author} {\bibfnamefont {J.-H.}\ \bibnamefont {Chu}},
  \bibinfo {author} {\bibfnamefont {J.}~\bibnamefont {Analytis}}, \bibinfo
  {author} {\bibfnamefont {Z.}~\bibnamefont {Liu}}, \bibinfo {author}
  {\bibfnamefont {K.}~\bibnamefont {Igarashi}}, \bibinfo {author}
  {\bibfnamefont {H.-H.}\ \bibnamefont {Kuo}}, \bibinfo {author} {\bibfnamefont
  {X.}~\bibnamefont {Qi}}, \bibinfo {author} {\bibfnamefont {S.-K.}\
  \bibnamefont {Mo}}, \bibinfo {author} {\bibfnamefont {R.}~\bibnamefont
  {Moore}}, \bibinfo {author} {\bibfnamefont {D.}~\bibnamefont {Lu}}, \emph
  {et~al.},\ }\bibfield  {title} {\bibinfo {title} {\textit{Massive Dirac
  fermion on the surface of a magnetically doped topological insulator}},\
  }\href {https://doi.org/10.1126/science.118992} {\bibfield  {journal}
  {\bibinfo  {journal} {Science}\ }\textbf {\bibinfo {volume} {329}},\ \bibinfo
  {pages} {659} (\bibinfo {year} {2010})}\BibitemShut {NoStop}%
\bibitem [{\citenamefont {Sessi}\ \emph {et~al.}(2016)\citenamefont {Sessi},
  \citenamefont {R{\"u}{\ss}mann}, \citenamefont {Bathon}, \citenamefont
  {Barla}, \citenamefont {Kokh}, \citenamefont {Tereshchenko}, \citenamefont
  {Fauth}, \citenamefont {Mahatha}, \citenamefont {Valbuena}, \citenamefont
  {Godey} \emph {et~al.}}]{sessi2016superparamagnetism}%
  \BibitemOpen
  \bibfield  {author} {\bibinfo {author} {\bibfnamefont {P.}~\bibnamefont
  {Sessi}}, \bibinfo {author} {\bibfnamefont {P.}~\bibnamefont
  {R{\"u}{\ss}mann}}, \bibinfo {author} {\bibfnamefont {T.}~\bibnamefont
  {Bathon}}, \bibinfo {author} {\bibfnamefont {A.}~\bibnamefont {Barla}},
  \bibinfo {author} {\bibfnamefont {K.}~\bibnamefont {Kokh}}, \bibinfo {author}
  {\bibfnamefont {O.}~\bibnamefont {Tereshchenko}}, \bibinfo {author}
  {\bibfnamefont {K.}~\bibnamefont {Fauth}}, \bibinfo {author} {\bibfnamefont
  {S.}~\bibnamefont {Mahatha}}, \bibinfo {author} {\bibfnamefont
  {M.}~\bibnamefont {Valbuena}}, \bibinfo {author} {\bibfnamefont
  {S.}~\bibnamefont {Godey}}, \emph {et~al.},\ }\bibfield  {title} {\bibinfo
  {title} {\textit{Superparamagnetism-induced mesoscopic electron focusing in
  topological insulators}},\ }\href
  {https://doi.org/10.1103/PhysRevB.94.075137} {\bibfield  {journal} {\bibinfo
  {journal} {Physical Review B}\ }\textbf {\bibinfo {volume} {94}},\ \bibinfo
  {pages} {075137} (\bibinfo {year} {2016})}\BibitemShut {NoStop}%
\bibitem [{\citenamefont {R{\"u}{\ss}mann}\ \emph {et~al.}(2018)\citenamefont
  {R{\"u}{\ss}mann}, \citenamefont {Mahatha}, \citenamefont {Sessi},
  \citenamefont {Valbuena}, \citenamefont {Bathon}, \citenamefont {Fauth},
  \citenamefont {Godey}, \citenamefont {Mugarza}, \citenamefont {Kokh},
  \citenamefont {Tereshchenko} \emph {et~al.}}]{russmann2018towards}%
  \BibitemOpen
  \bibfield  {author} {\bibinfo {author} {\bibfnamefont {P.}~\bibnamefont
  {R{\"u}{\ss}mann}}, \bibinfo {author} {\bibfnamefont {S.~K.}\ \bibnamefont
  {Mahatha}}, \bibinfo {author} {\bibfnamefont {P.}~\bibnamefont {Sessi}},
  \bibinfo {author} {\bibfnamefont {M.~A.}\ \bibnamefont {Valbuena}}, \bibinfo
  {author} {\bibfnamefont {T.}~\bibnamefont {Bathon}}, \bibinfo {author}
  {\bibfnamefont {K.}~\bibnamefont {Fauth}}, \bibinfo {author} {\bibfnamefont
  {S.}~\bibnamefont {Godey}}, \bibinfo {author} {\bibfnamefont
  {A.}~\bibnamefont {Mugarza}}, \bibinfo {author} {\bibfnamefont {K.~A.}\
  \bibnamefont {Kokh}}, \bibinfo {author} {\bibfnamefont {O.~E.}\ \bibnamefont
  {Tereshchenko}}, \emph {et~al.},\ }\bibfield  {title} {\bibinfo {title}
  {\textit{Towards microscopic control of the magnetic exchange coupling at the
  surface of a topological insulator}},\ }\href
  {https://doi.org/10.1088/2515-7639/aad02a DownloadArticle PDF} {\bibfield
  {journal} {\bibinfo  {journal} {Journal of Physics: Materials}\ }\textbf
  {\bibinfo {volume} {1}},\ \bibinfo {pages} {015002} (\bibinfo {year}
  {2018})}\BibitemShut {NoStop}%
\bibitem [{\citenamefont {Otrokov}\ \emph {et~al.}(2017)\citenamefont
  {Otrokov}, \citenamefont {Menshchikova}, \citenamefont {Rusinov},
  \citenamefont {Vergniory}, \citenamefont {Kuznetsov},\ and\ \citenamefont
  {Chulkov}}]{otrokov2017magnetic}%
  \BibitemOpen
  \bibfield  {author} {\bibinfo {author} {\bibfnamefont {M.~M.}\ \bibnamefont
  {Otrokov}}, \bibinfo {author} {\bibfnamefont {T.~V.}\ \bibnamefont
  {Menshchikova}}, \bibinfo {author} {\bibfnamefont {I.}~\bibnamefont
  {Rusinov}}, \bibinfo {author} {\bibfnamefont {M.}~\bibnamefont {Vergniory}},
  \bibinfo {author} {\bibfnamefont {V.~M.}\ \bibnamefont {Kuznetsov}},\ and\
  \bibinfo {author} {\bibfnamefont {E.~V.}\ \bibnamefont {Chulkov}},\
  }\bibfield  {title} {\bibinfo {title} {\textit{Magnetic extension as an
  efficient method for realizing the quantum anomalous hall state in
  topological insulators}},\ }\href {https://doi.org/10.1134/S0021364017050113}
  {\bibfield  {journal} {\bibinfo  {journal} {JETP Letters}\ }\textbf {\bibinfo
  {volume} {105}},\ \bibinfo {pages} {297} (\bibinfo {year}
  {2017})}\BibitemShut {NoStop}%
\bibitem [{\citenamefont {Kandala}\ \emph {et~al.}(2013)\citenamefont
  {Kandala}, \citenamefont {Richardella}, \citenamefont {Rench}, \citenamefont
  {Zhang}, \citenamefont {Flanagan},\ and\ \citenamefont
  {Samarth}}]{kandala2013growth}%
  \BibitemOpen
  \bibfield  {author} {\bibinfo {author} {\bibfnamefont {A.}~\bibnamefont
  {Kandala}}, \bibinfo {author} {\bibfnamefont {A.}~\bibnamefont
  {Richardella}}, \bibinfo {author} {\bibfnamefont {D.}~\bibnamefont {Rench}},
  \bibinfo {author} {\bibfnamefont {D.}~\bibnamefont {Zhang}}, \bibinfo
  {author} {\bibfnamefont {T.}~\bibnamefont {Flanagan}},\ and\ \bibinfo
  {author} {\bibfnamefont {N.}~\bibnamefont {Samarth}},\ }\bibfield  {title}
  {\bibinfo {title} {\textit{Growth and characterization of hybrid insulating
  ferromagnet-topological insulator heterostructure devices}},\ }\href
  {https://doi.org/10.1063/1.4831987} {\bibfield  {journal} {\bibinfo
  {journal} {Applied Physics Letters}\ }\textbf {\bibinfo {volume} {103}},\
  \bibinfo {pages} {202409} (\bibinfo {year} {2013})}\BibitemShut {NoStop}%
\bibitem [{\citenamefont {Hou}\ \emph {et~al.}(2019)\citenamefont {Hou},
  \citenamefont {Kim},\ and\ \citenamefont {Wu}}]{hou2019magnetizing}%
  \BibitemOpen
  \bibfield  {author} {\bibinfo {author} {\bibfnamefont {Y.}~\bibnamefont
  {Hou}}, \bibinfo {author} {\bibfnamefont {J.}~\bibnamefont {Kim}},\ and\
  \bibinfo {author} {\bibfnamefont {R.}~\bibnamefont {Wu}},\ }\bibfield
  {title} {\bibinfo {title} {\textit{Magnetizing topological surface states of
  Bi$_{2}$Se$_{3}$ with a CrI$_{3}$ monolayer}},\ }\href
  {https://doi.org/10.1126/sciadv.aaw1874} {\bibfield  {journal} {\bibinfo
  {journal} {Science advances}\ }\textbf {\bibinfo {volume} {5}},\ \bibinfo
  {pages} {eaaw1874} (\bibinfo {year} {2019})}\BibitemShut {NoStop}%
\bibitem [{\citenamefont {Chang}\ \emph
  {et~al.}(2013{\natexlab{a}})\citenamefont {Chang}, \citenamefont {Zhang},
  \citenamefont {Feng}, \citenamefont {Shen}, \citenamefont {Zhang},
  \citenamefont {Guo}, \citenamefont {Li}, \citenamefont {Ou}, \citenamefont
  {Wei}, \citenamefont {Wang} \emph {et~al.}}]{chang2013experimental}%
  \BibitemOpen
  \bibfield  {author} {\bibinfo {author} {\bibfnamefont {C.-Z.}\ \bibnamefont
  {Chang}}, \bibinfo {author} {\bibfnamefont {J.}~\bibnamefont {Zhang}},
  \bibinfo {author} {\bibfnamefont {X.}~\bibnamefont {Feng}}, \bibinfo {author}
  {\bibfnamefont {J.}~\bibnamefont {Shen}}, \bibinfo {author} {\bibfnamefont
  {Z.}~\bibnamefont {Zhang}}, \bibinfo {author} {\bibfnamefont
  {M.}~\bibnamefont {Guo}}, \bibinfo {author} {\bibfnamefont {K.}~\bibnamefont
  {Li}}, \bibinfo {author} {\bibfnamefont {Y.}~\bibnamefont {Ou}}, \bibinfo
  {author} {\bibfnamefont {P.}~\bibnamefont {Wei}}, \bibinfo {author}
  {\bibfnamefont {L.-L.}\ \bibnamefont {Wang}}, \emph {et~al.},\ }\bibfield
  {title} {\bibinfo {title} {\textit{Experimental observation of the quantum
  anomalous Hall effect in a magnetic topological insulator}},\ }\href
  {https://doi.org/10.1126/science.123441} {\bibfield  {journal} {\bibinfo
  {journal} {Science}\ }\textbf {\bibinfo {volume} {340}},\ \bibinfo {pages}
  {167} (\bibinfo {year} {2013}{\natexlab{a}})}\BibitemShut {NoStop}%
\bibitem [{\citenamefont {Chang}\ \emph {et~al.}(2015)\citenamefont {Chang},
  \citenamefont {Zhao}, \citenamefont {Kim}, \citenamefont {Zhang},
  \citenamefont {Assaf}, \citenamefont {Heiman}, \citenamefont {Zhang},
  \citenamefont {Liu}, \citenamefont {Chan},\ and\ \citenamefont
  {Moodera}}]{chang2015high}%
  \BibitemOpen
  \bibfield  {author} {\bibinfo {author} {\bibfnamefont {C.-Z.}\ \bibnamefont
  {Chang}}, \bibinfo {author} {\bibfnamefont {W.}~\bibnamefont {Zhao}},
  \bibinfo {author} {\bibfnamefont {D.~Y.}\ \bibnamefont {Kim}}, \bibinfo
  {author} {\bibfnamefont {H.}~\bibnamefont {Zhang}}, \bibinfo {author}
  {\bibfnamefont {B.~A.}\ \bibnamefont {Assaf}}, \bibinfo {author}
  {\bibfnamefont {D.}~\bibnamefont {Heiman}}, \bibinfo {author} {\bibfnamefont
  {S.-C.}\ \bibnamefont {Zhang}}, \bibinfo {author} {\bibfnamefont
  {C.}~\bibnamefont {Liu}}, \bibinfo {author} {\bibfnamefont {M.~H.}\
  \bibnamefont {Chan}},\ and\ \bibinfo {author} {\bibfnamefont {J.~S.}\
  \bibnamefont {Moodera}},\ }\bibfield  {title} {\bibinfo {title}
  {\textit{High-precision realization of robust quantum anomalous Hall state in
  a hard ferromagnetic topological insulator}},\ }\href
  {https://doi.org/doi.org/10.1038/nmat4204} {\bibfield  {journal} {\bibinfo
  {journal} {Nature Materials}\ }\textbf {\bibinfo {volume} {14}},\ \bibinfo
  {pages} {473} (\bibinfo {year} {2015})}\BibitemShut {NoStop}%
\bibitem [{\citenamefont {Otrokov}\ \emph {et~al.}(2019)\citenamefont
  {Otrokov}, \citenamefont {Klimovskikh}, \citenamefont {Bentmann},
  \citenamefont {Estyunin}, \citenamefont {Zeugner}, \citenamefont {Aliev},
  \citenamefont {Ga{\ss}}, \citenamefont {Wolter}, \citenamefont {Koroleva},
  \citenamefont {Shikin} \emph {et~al.}}]{otrokov2019prediction}%
  \BibitemOpen
  \bibfield  {author} {\bibinfo {author} {\bibfnamefont {M.~M.}\ \bibnamefont
  {Otrokov}}, \bibinfo {author} {\bibfnamefont {I.~I.}\ \bibnamefont
  {Klimovskikh}}, \bibinfo {author} {\bibfnamefont {H.}~\bibnamefont
  {Bentmann}}, \bibinfo {author} {\bibfnamefont {D.}~\bibnamefont {Estyunin}},
  \bibinfo {author} {\bibfnamefont {A.}~\bibnamefont {Zeugner}}, \bibinfo
  {author} {\bibfnamefont {Z.~S.}\ \bibnamefont {Aliev}}, \bibinfo {author}
  {\bibfnamefont {S.}~\bibnamefont {Ga{\ss}}}, \bibinfo {author} {\bibfnamefont
  {A.}~\bibnamefont {Wolter}}, \bibinfo {author} {\bibfnamefont
  {A.}~\bibnamefont {Koroleva}}, \bibinfo {author} {\bibfnamefont {A.~M.}\
  \bibnamefont {Shikin}}, \emph {et~al.},\ }\bibfield  {title} {\bibinfo
  {title} {\textit{Prediction and observation of an antiferromagnetic
  topological insulator}},\ }\href {https://doi.org/10.1038/s41586-019-1840-9}
  {\bibfield  {journal} {\bibinfo  {journal} {Nature}\ }\textbf {\bibinfo
  {volume} {576}},\ \bibinfo {pages} {416} (\bibinfo {year}
  {2019})}\BibitemShut {NoStop}%
\bibitem [{\citenamefont {Rienks}\ \emph {et~al.}(2019)\citenamefont {Rienks},
  \citenamefont {Wimmer}, \citenamefont {S{\'a}nchez-Barriga}, \citenamefont
  {Caha}, \citenamefont {Mandal}, \citenamefont {R\r{u}\v{z}i\v{c}ka},
  \citenamefont {Ney}, \citenamefont {Steiner}, \citenamefont {Volobuev},
  \citenamefont {Groi{\ss}} \emph {et~al.}}]{rienks2019large}%
  \BibitemOpen
  \bibfield  {author} {\bibinfo {author} {\bibfnamefont {E.~D.}\ \bibnamefont
  {Rienks}}, \bibinfo {author} {\bibfnamefont {S.}~\bibnamefont {Wimmer}},
  \bibinfo {author} {\bibfnamefont {J.}~\bibnamefont {S{\'a}nchez-Barriga}},
  \bibinfo {author} {\bibfnamefont {O.}~\bibnamefont {Caha}}, \bibinfo {author}
  {\bibfnamefont {P.~S.}\ \bibnamefont {Mandal}}, \bibinfo {author}
  {\bibfnamefont {J.}~\bibnamefont {R\r{u}\v{z}i\v{c}ka}}, \bibinfo {author}
  {\bibfnamefont {A.}~\bibnamefont {Ney}}, \bibinfo {author} {\bibfnamefont
  {H.}~\bibnamefont {Steiner}}, \bibinfo {author} {\bibfnamefont {V.~V.}\
  \bibnamefont {Volobuev}}, \bibinfo {author} {\bibfnamefont {H.}~\bibnamefont
  {Groi{\ss}}}, \emph {et~al.},\ }\bibfield  {title} {\bibinfo {title}
  {\textit{Large magnetic gap at the Dirac point in
  Bi$_{2}$Te$_{3}$/MnBi$_{2}$Te$_{4}$ heterostructures}},\ }\href
  {https://doi.org/10.1038/s41586-019-1826-7} {\bibfield  {journal} {\bibinfo
  {journal} {Nature}\ }\textbf {\bibinfo {volume} {576}},\ \bibinfo {pages}
  {423} (\bibinfo {year} {2019})}\BibitemShut {NoStop}%
\bibitem [{\citenamefont {Garnica}\ \emph {et~al.}(2022)\citenamefont
  {Garnica}, \citenamefont {Otrokov}, \citenamefont {Aguilar}, \citenamefont
  {Klimovskikh}, \citenamefont {Estyunin}, \citenamefont {Aliev}, \citenamefont
  {Amiraslanov}, \citenamefont {Abdullayev}, \citenamefont {Zverev},
  \citenamefont {Babanly} \emph {et~al.}}]{garnica2022native}%
  \BibitemOpen
  \bibfield  {author} {\bibinfo {author} {\bibfnamefont {M.}~\bibnamefont
  {Garnica}}, \bibinfo {author} {\bibfnamefont {M.~M.}\ \bibnamefont
  {Otrokov}}, \bibinfo {author} {\bibfnamefont {P.~C.}\ \bibnamefont
  {Aguilar}}, \bibinfo {author} {\bibfnamefont {I.}~\bibnamefont
  {Klimovskikh}}, \bibinfo {author} {\bibfnamefont {D.}~\bibnamefont
  {Estyunin}}, \bibinfo {author} {\bibfnamefont {Z.~S.}\ \bibnamefont {Aliev}},
  \bibinfo {author} {\bibfnamefont {I.~R.}\ \bibnamefont {Amiraslanov}},
  \bibinfo {author} {\bibfnamefont {N.~A.}\ \bibnamefont {Abdullayev}},
  \bibinfo {author} {\bibfnamefont {V.~N.}\ \bibnamefont {Zverev}}, \bibinfo
  {author} {\bibfnamefont {M.~B.}\ \bibnamefont {Babanly}}, \emph {et~al.},\
  }\bibfield  {title} {\bibinfo {title} {\textit{Native point defects and their
  implications for the Dirac point gap at MnBi2Te4 (0001)}},\ }\href
  {https://doi.org/10.1038/s41535-021-00414-6} {\bibfield  {journal} {\bibinfo
  {journal} {npj Quantum Materials}\ }\textbf {\bibinfo {volume} {7}},\
  \bibinfo {pages} {1} (\bibinfo {year} {2022})}\BibitemShut {NoStop}%
\bibitem [{\citenamefont {Grauer}\ \emph {et~al.}(2015)\citenamefont {Grauer},
  \citenamefont {Schreyeck}, \citenamefont {Winnerlein}, \citenamefont
  {Brunner}, \citenamefont {Gould},\ and\ \citenamefont
  {Molenkamp}}]{grauer2015coincidence}%
  \BibitemOpen
  \bibfield  {author} {\bibinfo {author} {\bibfnamefont {S.}~\bibnamefont
  {Grauer}}, \bibinfo {author} {\bibfnamefont {S.}~\bibnamefont {Schreyeck}},
  \bibinfo {author} {\bibfnamefont {M.}~\bibnamefont {Winnerlein}}, \bibinfo
  {author} {\bibfnamefont {K.}~\bibnamefont {Brunner}}, \bibinfo {author}
  {\bibfnamefont {C.}~\bibnamefont {Gould}},\ and\ \bibinfo {author}
  {\bibfnamefont {L.}~\bibnamefont {Molenkamp}},\ }\bibfield  {title} {\bibinfo
  {title} {\textit{Coincidence of superparamagnetism and perfect quantization
  in the quantum anomalous Hall state}},\ }\href
  {https://doi.org/10.1103/PhysRevB.92.201304} {\bibfield  {journal} {\bibinfo
  {journal} {Physical Review B}\ }\textbf {\bibinfo {volume} {92}},\ \bibinfo
  {pages} {201304} (\bibinfo {year} {2015})}\BibitemShut {NoStop}%
\bibitem [{\citenamefont {Pan}\ \emph {et~al.}(2020)\citenamefont {Pan},
  \citenamefont {Liu}, \citenamefont {He}, \citenamefont {Stern}, \citenamefont
  {Yin}, \citenamefont {Che}, \citenamefont {Shao}, \citenamefont {Zhang},
  \citenamefont {Deng}, \citenamefont {Yang} \emph {et~al.}}]{pan2020probing}%
  \BibitemOpen
  \bibfield  {author} {\bibinfo {author} {\bibfnamefont {L.}~\bibnamefont
  {Pan}}, \bibinfo {author} {\bibfnamefont {X.}~\bibnamefont {Liu}}, \bibinfo
  {author} {\bibfnamefont {Q.~L.}\ \bibnamefont {He}}, \bibinfo {author}
  {\bibfnamefont {A.}~\bibnamefont {Stern}}, \bibinfo {author} {\bibfnamefont
  {G.}~\bibnamefont {Yin}}, \bibinfo {author} {\bibfnamefont {X.}~\bibnamefont
  {Che}}, \bibinfo {author} {\bibfnamefont {Q.}~\bibnamefont {Shao}}, \bibinfo
  {author} {\bibfnamefont {P.}~\bibnamefont {Zhang}}, \bibinfo {author}
  {\bibfnamefont {P.}~\bibnamefont {Deng}}, \bibinfo {author} {\bibfnamefont
  {C.-Y.}\ \bibnamefont {Yang}}, \emph {et~al.},\ }\bibfield  {title} {\bibinfo
  {title} {\textit{Probing the low-temperature limit of the quantum anomalous
  Hall effect}},\ }\href {https://doi.org/10.1126/sciadv.aaz35} {\bibfield
  {journal} {\bibinfo  {journal} {Science advances}\ }\textbf {\bibinfo
  {volume} {6}},\ \bibinfo {pages} {eaaz3595} (\bibinfo {year}
  {2020})}\BibitemShut {NoStop}%
\bibitem [{\citenamefont {Eelbo}\ \emph {et~al.}(2014)\citenamefont {Eelbo},
  \citenamefont {Wa{\'s}niowska}, \citenamefont {Sikora}, \citenamefont
  {Dobrza{\'n}ski}, \citenamefont {Koz{\l}owski}, \citenamefont {Pulkin},
  \citenamefont {Aut{\`e}s}, \citenamefont {Miotkowski}, \citenamefont
  {Yazyev},\ and\ \citenamefont {Wiesendanger}}]{eelbo2014strong}%
  \BibitemOpen
  \bibfield  {author} {\bibinfo {author} {\bibfnamefont {T.}~\bibnamefont
  {Eelbo}}, \bibinfo {author} {\bibfnamefont {M.}~\bibnamefont
  {Wa{\'s}niowska}}, \bibinfo {author} {\bibfnamefont {M.}~\bibnamefont
  {Sikora}}, \bibinfo {author} {\bibfnamefont {M.}~\bibnamefont
  {Dobrza{\'n}ski}}, \bibinfo {author} {\bibfnamefont {A.}~\bibnamefont
  {Koz{\l}owski}}, \bibinfo {author} {\bibfnamefont {A.}~\bibnamefont
  {Pulkin}}, \bibinfo {author} {\bibfnamefont {G.}~\bibnamefont {Aut{\`e}s}},
  \bibinfo {author} {\bibfnamefont {I.}~\bibnamefont {Miotkowski}}, \bibinfo
  {author} {\bibfnamefont {O.}~\bibnamefont {Yazyev}},\ and\ \bibinfo {author}
  {\bibfnamefont {R.}~\bibnamefont {Wiesendanger}},\ }\bibfield  {title}
  {\bibinfo {title} {\textit{Strong out-of-plane magnetic anisotropy of Fe
  adatoms on Bi$_{2}$Te$_{3}$}},\ }\href
  {https://doi.org/10.1103/PhysRevB.89.104424} {\bibfield  {journal} {\bibinfo
  {journal} {Physical Review B}\ }\textbf {\bibinfo {volume} {89}},\ \bibinfo
  {pages} {104424} (\bibinfo {year} {2014})}\BibitemShut {NoStop}%
\bibitem [{\citenamefont {Gambardella}\ \emph {et~al.}(2003)\citenamefont
  {Gambardella}, \citenamefont {Rusponi}, \citenamefont {Veronese},
  \citenamefont {Dhesi}, \citenamefont {Grazioli}, \citenamefont {Dallmeyer},
  \citenamefont {Cabria}, \citenamefont {Zeller}, \citenamefont {Dederichs},
  \citenamefont {Kern} \emph {et~al.}}]{gambardella2003giant}%
  \BibitemOpen
  \bibfield  {author} {\bibinfo {author} {\bibfnamefont {P.}~\bibnamefont
  {Gambardella}}, \bibinfo {author} {\bibfnamefont {S.}~\bibnamefont
  {Rusponi}}, \bibinfo {author} {\bibfnamefont {M.}~\bibnamefont {Veronese}},
  \bibinfo {author} {\bibfnamefont {S.}~\bibnamefont {Dhesi}}, \bibinfo
  {author} {\bibfnamefont {C.}~\bibnamefont {Grazioli}}, \bibinfo {author}
  {\bibfnamefont {A.}~\bibnamefont {Dallmeyer}}, \bibinfo {author}
  {\bibfnamefont {I.}~\bibnamefont {Cabria}}, \bibinfo {author} {\bibfnamefont
  {R.}~\bibnamefont {Zeller}}, \bibinfo {author} {\bibfnamefont
  {P.}~\bibnamefont {Dederichs}}, \bibinfo {author} {\bibfnamefont
  {K.}~\bibnamefont {Kern}}, \emph {et~al.},\ }\bibfield  {title} {\bibinfo
  {title} {\textit{Giant magnetic anisotropy of single cobalt atoms and
  nanoparticles}},\ }\href {https://doi.org/10.1126/science.1082857} {\bibfield
   {journal} {\bibinfo  {journal} {Science}\ }\textbf {\bibinfo {volume}
  {300}},\ \bibinfo {pages} {1130} (\bibinfo {year} {2003})}\BibitemShut
  {NoStop}%
\bibitem [{\citenamefont {Wray}\ \emph {et~al.}(2011)\citenamefont {Wray},
  \citenamefont {Xu}, \citenamefont {Xia}, \citenamefont {Hsieh}, \citenamefont
  {Fedorov}, \citenamefont {Hor}, \citenamefont {Cava}, \citenamefont {Bansil},
  \citenamefont {Lin},\ and\ \citenamefont {Hasan}}]{wray2011topological}%
  \BibitemOpen
  \bibfield  {author} {\bibinfo {author} {\bibfnamefont {L.~A.}\ \bibnamefont
  {Wray}}, \bibinfo {author} {\bibfnamefont {S.-Y.}\ \bibnamefont {Xu}},
  \bibinfo {author} {\bibfnamefont {Y.}~\bibnamefont {Xia}}, \bibinfo {author}
  {\bibfnamefont {D.}~\bibnamefont {Hsieh}}, \bibinfo {author} {\bibfnamefont
  {A.~V.}\ \bibnamefont {Fedorov}}, \bibinfo {author} {\bibfnamefont {Y.~S.}\
  \bibnamefont {Hor}}, \bibinfo {author} {\bibfnamefont {R.~J.}\ \bibnamefont
  {Cava}}, \bibinfo {author} {\bibfnamefont {A.}~\bibnamefont {Bansil}},
  \bibinfo {author} {\bibfnamefont {H.}~\bibnamefont {Lin}},\ and\ \bibinfo
  {author} {\bibfnamefont {M.~Z.}\ \bibnamefont {Hasan}},\ }\bibfield  {title}
  {\bibinfo {title} {\textit{A topological insulator surface under strong
  Coulomb, magnetic and disorder perturbations}},\ }\href
  {https://doi.org/10.1038/nphys1838} {\bibfield  {journal} {\bibinfo
  {journal} {Nature Physics}\ }\textbf {\bibinfo {volume} {7}},\ \bibinfo
  {pages} {32} (\bibinfo {year} {2011})}\BibitemShut {NoStop}%
\bibitem [{\citenamefont {Scholz}\ \emph {et~al.}(2012)\citenamefont {Scholz},
  \citenamefont {S{\'a}nchez-Barriga}, \citenamefont {Marchenko}, \citenamefont
  {Varykhalov}, \citenamefont {Volykhov}, \citenamefont {Yashina},\ and\
  \citenamefont {Rader}}]{scholz2012tolerance}%
  \BibitemOpen
  \bibfield  {author} {\bibinfo {author} {\bibfnamefont {M.}~\bibnamefont
  {Scholz}}, \bibinfo {author} {\bibfnamefont {J.}~\bibnamefont
  {S{\'a}nchez-Barriga}}, \bibinfo {author} {\bibfnamefont {D.}~\bibnamefont
  {Marchenko}}, \bibinfo {author} {\bibfnamefont {A.}~\bibnamefont
  {Varykhalov}}, \bibinfo {author} {\bibfnamefont {A.}~\bibnamefont
  {Volykhov}}, \bibinfo {author} {\bibfnamefont {L.}~\bibnamefont {Yashina}},\
  and\ \bibinfo {author} {\bibfnamefont {O.}~\bibnamefont {Rader}},\ }\bibfield
   {title} {\bibinfo {title} {\textit{Tolerance of topological surface states
  towards magnetic moments: Fe on Bi$_{2}$Se$_{3}$}},\ }\href
  {https://doi.org/10.1103/PhysRevLett.108.256810} {\bibfield  {journal}
  {\bibinfo  {journal} {Physical Review Letters}\ }\textbf {\bibinfo {volume}
  {108}},\ \bibinfo {pages} {256810} (\bibinfo {year} {2012})}\BibitemShut
  {NoStop}%
\bibitem [{\citenamefont {Valla}\ \emph {et~al.}(2012)\citenamefont {Valla},
  \citenamefont {Pan}, \citenamefont {Gardner}, \citenamefont {Lee},\ and\
  \citenamefont {Chu}}]{valla2012photoemission}%
  \BibitemOpen
  \bibfield  {author} {\bibinfo {author} {\bibfnamefont {T.}~\bibnamefont
  {Valla}}, \bibinfo {author} {\bibfnamefont {Z.-H.}\ \bibnamefont {Pan}},
  \bibinfo {author} {\bibfnamefont {D.}~\bibnamefont {Gardner}}, \bibinfo
  {author} {\bibfnamefont {Y.}~\bibnamefont {Lee}},\ and\ \bibinfo {author}
  {\bibfnamefont {S.}~\bibnamefont {Chu}},\ }\bibfield  {title} {\bibinfo
  {title} {\textit{Photoemission spectroscopy of magnetic and nonmagnetic
  impurities on the surface of the Bi$_{2}$Se$_{3}$ topological insulator}},\
  }\href {https://doi.org/10.1103/PhysRevLett.108.117601} {\bibfield  {journal}
  {\bibinfo  {journal} {Physical Review Letters}\ }\textbf {\bibinfo {volume}
  {108}},\ \bibinfo {pages} {117601} (\bibinfo {year} {2012})}\BibitemShut
  {NoStop}%
\bibitem [{\citenamefont {Honolka}\ \emph {et~al.}(2012)\citenamefont
  {Honolka}, \citenamefont {Khajetoorians}, \citenamefont {Sessi},
  \citenamefont {Wehling}, \citenamefont {Stepanow}, \citenamefont {Mi},
  \citenamefont {Iversen}, \citenamefont {Schlenk}, \citenamefont {Wiebe},
  \citenamefont {Brookes} \emph {et~al.}}]{honolka2012plane}%
  \BibitemOpen
  \bibfield  {author} {\bibinfo {author} {\bibfnamefont {J.}~\bibnamefont
  {Honolka}}, \bibinfo {author} {\bibfnamefont {A.}~\bibnamefont
  {Khajetoorians}}, \bibinfo {author} {\bibfnamefont {V.}~\bibnamefont
  {Sessi}}, \bibinfo {author} {\bibfnamefont {T.}~\bibnamefont {Wehling}},
  \bibinfo {author} {\bibfnamefont {S.}~\bibnamefont {Stepanow}}, \bibinfo
  {author} {\bibfnamefont {J.-L.}\ \bibnamefont {Mi}}, \bibinfo {author}
  {\bibfnamefont {B.}~\bibnamefont {Iversen}}, \bibinfo {author} {\bibfnamefont
  {T.}~\bibnamefont {Schlenk}}, \bibinfo {author} {\bibfnamefont
  {J.}~\bibnamefont {Wiebe}}, \bibinfo {author} {\bibfnamefont
  {N.}~\bibnamefont {Brookes}}, \emph {et~al.},\ }\bibfield  {title} {\bibinfo
  {title} {\textit{In-plane magnetic anisotropy of Fe atoms on
  Bi$_{2}$Se$_{3}$(111)}},\ }\href
  {https://doi.org/10.1103/PhysRevLett.108.256811} {\bibfield  {journal}
  {\bibinfo  {journal} {Physical Review Letters}\ }\textbf {\bibinfo {volume}
  {108}},\ \bibinfo {pages} {256811} (\bibinfo {year} {2012})}\BibitemShut
  {NoStop}%
\bibitem [{\citenamefont {Beidenkopf}\ \emph {et~al.}(2011)\citenamefont
  {Beidenkopf}, \citenamefont {Roushan}, \citenamefont {Seo}, \citenamefont
  {Gorman}, \citenamefont {Drozdov}, \citenamefont {Hor}, \citenamefont
  {Cava},\ and\ \citenamefont {Yazdani}}]{beidenkopf2011spatial}%
  \BibitemOpen
  \bibfield  {author} {\bibinfo {author} {\bibfnamefont {H.}~\bibnamefont
  {Beidenkopf}}, \bibinfo {author} {\bibfnamefont {P.}~\bibnamefont {Roushan}},
  \bibinfo {author} {\bibfnamefont {J.}~\bibnamefont {Seo}}, \bibinfo {author}
  {\bibfnamefont {L.}~\bibnamefont {Gorman}}, \bibinfo {author} {\bibfnamefont
  {I.}~\bibnamefont {Drozdov}}, \bibinfo {author} {\bibfnamefont {Y.~S.}\
  \bibnamefont {Hor}}, \bibinfo {author} {\bibfnamefont {R.~J.}\ \bibnamefont
  {Cava}},\ and\ \bibinfo {author} {\bibfnamefont {A.}~\bibnamefont
  {Yazdani}},\ }\bibfield  {title} {\bibinfo {title} {\textit{Spatial
  fluctuations of helical Dirac fermions on the surface of topological
  insulators}},\ }\href {https://doi.org/10.1038/nphys2108} {\bibfield
  {journal} {\bibinfo  {journal} {Nature Physics}\ }\textbf {\bibinfo {volume}
  {7}},\ \bibinfo {pages} {939} (\bibinfo {year} {2011})}\BibitemShut {NoStop}%
\bibitem [{\citenamefont {Chang}\ \emph
  {et~al.}(2013{\natexlab{b}})\citenamefont {Chang}, \citenamefont {Zhang},
  \citenamefont {Liu}, \citenamefont {Zhang}, \citenamefont {Feng},
  \citenamefont {Li}, \citenamefont {Wang}, \citenamefont {Chen}, \citenamefont
  {Dai}, \citenamefont {Fang} \emph {et~al.}}]{chang2013thin}%
  \BibitemOpen
  \bibfield  {author} {\bibinfo {author} {\bibfnamefont {C.-Z.}\ \bibnamefont
  {Chang}}, \bibinfo {author} {\bibfnamefont {J.}~\bibnamefont {Zhang}},
  \bibinfo {author} {\bibfnamefont {M.}~\bibnamefont {Liu}}, \bibinfo {author}
  {\bibfnamefont {Z.}~\bibnamefont {Zhang}}, \bibinfo {author} {\bibfnamefont
  {X.}~\bibnamefont {Feng}}, \bibinfo {author} {\bibfnamefont {K.}~\bibnamefont
  {Li}}, \bibinfo {author} {\bibfnamefont {L.-L.}\ \bibnamefont {Wang}},
  \bibinfo {author} {\bibfnamefont {X.}~\bibnamefont {Chen}}, \bibinfo {author}
  {\bibfnamefont {X.}~\bibnamefont {Dai}}, \bibinfo {author} {\bibfnamefont
  {Z.}~\bibnamefont {Fang}}, \emph {et~al.},\ }\bibfield  {title} {\bibinfo
  {title} {\textit{Thin films of magnetically doped topological insulator with
  carrier-independent long-range ferromagnetic order}},\ }\href
  {https://doi.org/10.1002/adma.201203493} {\bibfield  {journal} {\bibinfo
  {journal} {Advanced materials}\ }\textbf {\bibinfo {volume} {25}},\ \bibinfo
  {pages} {1065} (\bibinfo {year} {2013}{\natexlab{b}})}\BibitemShut {NoStop}%
\bibitem [{\citenamefont {Zhang}\ \emph {et~al.}(2013)\citenamefont {Zhang},
  \citenamefont {Chang}, \citenamefont {Tang}, \citenamefont {Zhang},
  \citenamefont {Feng}, \citenamefont {Li}, \citenamefont {Wang}, \citenamefont
  {Chen}, \citenamefont {Liu}, \citenamefont {Duan} \emph
  {et~al.}}]{zhang2013topology}%
  \BibitemOpen
  \bibfield  {author} {\bibinfo {author} {\bibfnamefont {J.}~\bibnamefont
  {Zhang}}, \bibinfo {author} {\bibfnamefont {C.-Z.}\ \bibnamefont {Chang}},
  \bibinfo {author} {\bibfnamefont {P.}~\bibnamefont {Tang}}, \bibinfo {author}
  {\bibfnamefont {Z.}~\bibnamefont {Zhang}}, \bibinfo {author} {\bibfnamefont
  {X.}~\bibnamefont {Feng}}, \bibinfo {author} {\bibfnamefont {K.}~\bibnamefont
  {Li}}, \bibinfo {author} {\bibfnamefont {L.-l.}\ \bibnamefont {Wang}},
  \bibinfo {author} {\bibfnamefont {X.}~\bibnamefont {Chen}}, \bibinfo {author}
  {\bibfnamefont {C.}~\bibnamefont {Liu}}, \bibinfo {author} {\bibfnamefont
  {W.}~\bibnamefont {Duan}}, \emph {et~al.},\ }\bibfield  {title} {\bibinfo
  {title} {\textit{Topology-driven magnetic quantum phase transition in
  topological insulators}},\ }\href {https://doi.org/10.1126/science.1230905}
  {\bibfield  {journal} {\bibinfo  {journal} {Science}\ }\textbf {\bibinfo
  {volume} {339}},\ \bibinfo {pages} {1582} (\bibinfo {year}
  {2013})}\BibitemShut {NoStop}%
\bibitem [{\citenamefont {Vobornik}\ \emph {et~al.}(2014)\citenamefont
  {Vobornik}, \citenamefont {Panaccione}, \citenamefont {Fujii}, \citenamefont
  {Zhu}, \citenamefont {Offi}, \citenamefont {Salles}, \citenamefont
  {Borgatti}, \citenamefont {Torelli}, \citenamefont {Rueff}, \citenamefont
  {Ceolin} \emph {et~al.}}]{vobornik2014observation}%
  \BibitemOpen
  \bibfield  {author} {\bibinfo {author} {\bibfnamefont {I.}~\bibnamefont
  {Vobornik}}, \bibinfo {author} {\bibfnamefont {G.}~\bibnamefont
  {Panaccione}}, \bibinfo {author} {\bibfnamefont {J.}~\bibnamefont {Fujii}},
  \bibinfo {author} {\bibfnamefont {Z.-H.}\ \bibnamefont {Zhu}}, \bibinfo
  {author} {\bibfnamefont {F.}~\bibnamefont {Offi}}, \bibinfo {author}
  {\bibfnamefont {B.~R.}\ \bibnamefont {Salles}}, \bibinfo {author}
  {\bibfnamefont {F.}~\bibnamefont {Borgatti}}, \bibinfo {author}
  {\bibfnamefont {P.}~\bibnamefont {Torelli}}, \bibinfo {author} {\bibfnamefont
  {J.~P.}\ \bibnamefont {Rueff}}, \bibinfo {author} {\bibfnamefont
  {D.}~\bibnamefont {Ceolin}}, \emph {et~al.},\ }\bibfield  {title} {\bibinfo
  {title} {\textit{Observation of distinct bulk and surface chemical
  environments in a topological insulator under magnetic doping}},\ }\href
  {https://doi.org/10.1021/jp502729u} {\bibfield  {journal} {\bibinfo
  {journal} {The Journal of Physical Chemistry C}\ }\textbf {\bibinfo {volume}
  {118}},\ \bibinfo {pages} {12333} (\bibinfo {year} {2014})}\BibitemShut
  {NoStop}%
\bibitem [{\citenamefont {Cuxart}\ \emph {et~al.}(2020)\citenamefont {Cuxart},
  \citenamefont {Valbuena}, \citenamefont {Robles}, \citenamefont {Moreno},
  \citenamefont {Bonell}, \citenamefont {Sauthier}, \citenamefont {Imaz},
  \citenamefont {Xu}, \citenamefont {Nistor}, \citenamefont {Barla} \emph
  {et~al.}}]{cuxart2020molecular}%
  \BibitemOpen
  \bibfield  {author} {\bibinfo {author} {\bibfnamefont {M.~G.}\ \bibnamefont
  {Cuxart}}, \bibinfo {author} {\bibfnamefont {M.~A.}\ \bibnamefont
  {Valbuena}}, \bibinfo {author} {\bibfnamefont {R.}~\bibnamefont {Robles}},
  \bibinfo {author} {\bibfnamefont {C.}~\bibnamefont {Moreno}}, \bibinfo
  {author} {\bibfnamefont {F.}~\bibnamefont {Bonell}}, \bibinfo {author}
  {\bibfnamefont {G.}~\bibnamefont {Sauthier}}, \bibinfo {author}
  {\bibfnamefont {I.}~\bibnamefont {Imaz}}, \bibinfo {author} {\bibfnamefont
  {H.}~\bibnamefont {Xu}}, \bibinfo {author} {\bibfnamefont {C.}~\bibnamefont
  {Nistor}}, \bibinfo {author} {\bibfnamefont {A.}~\bibnamefont {Barla}}, \emph
  {et~al.},\ }\bibfield  {title} {\bibinfo {title} {\textit{Molecular approach
  for engineering interfacial interactions in magnetic/topological insulator
  heterostructures}},\ }\href {https://doi.org/10.1021/acsnano.0c02498}
  {\bibfield  {journal} {\bibinfo  {journal} {ACS Nano}\ }\textbf {\bibinfo
  {volume} {14}},\ \bibinfo {pages} {6285} (\bibinfo {year}
  {2020})}\BibitemShut {NoStop}%
\bibitem [{\citenamefont {Abdalla}\ \emph {et~al.}(2013)\citenamefont
  {Abdalla}, \citenamefont {Seixas}, \citenamefont {Schmidt}, \citenamefont
  {Miwa},\ and\ \citenamefont {Fazzio}}]{abdalla2013topological}%
  \BibitemOpen
  \bibfield  {author} {\bibinfo {author} {\bibfnamefont {L.~B.}\ \bibnamefont
  {Abdalla}}, \bibinfo {author} {\bibfnamefont {L.}~\bibnamefont {Seixas}},
  \bibinfo {author} {\bibfnamefont {T.}~\bibnamefont {Schmidt}}, \bibinfo
  {author} {\bibfnamefont {R.}~\bibnamefont {Miwa}},\ and\ \bibinfo {author}
  {\bibfnamefont {A.}~\bibnamefont {Fazzio}},\ }\bibfield  {title} {\bibinfo
  {title} {\textit{Topological insulator Bi$_{2}$Se$_{3}$(111) surface doped
  with transition metals: An ab initio investigation}},\ }\href
  {https://doi.org/10.1103/PhysRevB.88.045312} {\bibfield  {journal} {\bibinfo
  {journal} {Physical Review B}\ }\textbf {\bibinfo {volume} {88}},\ \bibinfo
  {pages} {045312} (\bibinfo {year} {2013})}\BibitemShut {NoStop}%
\bibitem [{\citenamefont {Jensen}\ and\ \citenamefont
  {Mackintosh}(1991)}]{jensen1991rare}%
  \BibitemOpen
  \bibfield  {author} {\bibinfo {author} {\bibfnamefont {J.}~\bibnamefont
  {Jensen}}\ and\ \bibinfo {author} {\bibfnamefont {A.~R.}\ \bibnamefont
  {Mackintosh}},\ }\href {https://www.fys.ku.dk/~jjensen/Book/Ebook.pdf} {\emph
  {\bibinfo {title} {\textit{Rare earth magnetism}}}}\ (\bibinfo  {publisher}
  {Clarendon Press Oxford},\ \bibinfo {year} {1991})\BibitemShut {NoStop}%
\bibitem [{\citenamefont {Harrison}\ \emph
  {et~al.}(2015{\natexlab{a}})\citenamefont {Harrison}, \citenamefont
  {Collins-McIntyre}, \citenamefont {Zhang}, \citenamefont {Baker},
  \citenamefont {Figueroa}, \citenamefont {Kellock}, \citenamefont {Pushp},
  \citenamefont {Parkin}, \citenamefont {Harris}, \citenamefont {Van Der~Laan}
  \emph {et~al.}}]{harrison2015study}%
  \BibitemOpen
  \bibfield  {author} {\bibinfo {author} {\bibfnamefont {S.}~\bibnamefont
  {Harrison}}, \bibinfo {author} {\bibfnamefont {L.}~\bibnamefont
  {Collins-McIntyre}}, \bibinfo {author} {\bibfnamefont {S.}~\bibnamefont
  {Zhang}}, \bibinfo {author} {\bibfnamefont {A.}~\bibnamefont {Baker}},
  \bibinfo {author} {\bibfnamefont {A.}~\bibnamefont {Figueroa}}, \bibinfo
  {author} {\bibfnamefont {A.}~\bibnamefont {Kellock}}, \bibinfo {author}
  {\bibfnamefont {A.}~\bibnamefont {Pushp}}, \bibinfo {author} {\bibfnamefont
  {S.}~\bibnamefont {Parkin}}, \bibinfo {author} {\bibfnamefont
  {J.}~\bibnamefont {Harris}}, \bibinfo {author} {\bibfnamefont
  {G.}~\bibnamefont {Van Der~Laan}}, \emph {et~al.},\ }\bibfield  {title}
  {\bibinfo {title} {\textit{Study of Dy-doped Bi$_{2}$Te$_{3}$: thin film
  growth and magnetic properties}},\ }\href
  {https://doi.org/10.1088/0953-8984/27/24/245602} {\bibfield  {journal}
  {\bibinfo  {journal} {Journal of Physics: Condensed Matter}\ }\textbf
  {\bibinfo {volume} {27}},\ \bibinfo {pages} {245602} (\bibinfo {year}
  {2015}{\natexlab{a}})}\BibitemShut {NoStop}%
\bibitem [{\citenamefont {Ormaza}\ \emph {et~al.}(2016)\citenamefont {Ormaza},
  \citenamefont {Fern{\'a}ndez}, \citenamefont {Ilyn}, \citenamefont {Magana},
  \citenamefont {Xu}, \citenamefont {Verstraete}, \citenamefont {Gastaldo},
  \citenamefont {Valbuena}, \citenamefont {Gargiani}, \citenamefont {Mugarza}
  \emph {et~al.}}]{ormaza2016high}%
  \BibitemOpen
  \bibfield  {author} {\bibinfo {author} {\bibfnamefont {M.}~\bibnamefont
  {Ormaza}}, \bibinfo {author} {\bibfnamefont {L.}~\bibnamefont
  {Fern{\'a}ndez}}, \bibinfo {author} {\bibfnamefont {M.}~\bibnamefont {Ilyn}},
  \bibinfo {author} {\bibfnamefont {A.}~\bibnamefont {Magana}}, \bibinfo
  {author} {\bibfnamefont {B.}~\bibnamefont {Xu}}, \bibinfo {author}
  {\bibfnamefont {M.}~\bibnamefont {Verstraete}}, \bibinfo {author}
  {\bibfnamefont {M.}~\bibnamefont {Gastaldo}}, \bibinfo {author}
  {\bibfnamefont {M.}~\bibnamefont {Valbuena}}, \bibinfo {author}
  {\bibfnamefont {P.}~\bibnamefont {Gargiani}}, \bibinfo {author}
  {\bibfnamefont {A.}~\bibnamefont {Mugarza}}, \emph {et~al.},\ }\bibfield
  {title} {\bibinfo {title} {\textit{High temperature ferromagnetism in a
  GdAg$_{2}$ monolayer}},\ }\href
  {https://doi.org/10.1021/acs.nanolett.6b01197} {\bibfield  {journal}
  {\bibinfo  {journal} {Nano Letters}\ }\textbf {\bibinfo {volume} {16}},\
  \bibinfo {pages} {4230} (\bibinfo {year} {2016})}\BibitemShut {NoStop}%
\bibitem [{\citenamefont {Fern{\'a}ndez}\ \emph {et~al.}(2020)\citenamefont
  {Fern{\'a}ndez}, \citenamefont {Blanco-Rey}, \citenamefont
  {Castrillo-Bodero}, \citenamefont {Ilyn}, \citenamefont {Ali}, \citenamefont
  {Turco}, \citenamefont {Corso}, \citenamefont {Ormaza}, \citenamefont
  {Gargiani}, \citenamefont {Valbuena} \emph
  {et~al.}}]{fernandez2020influence}%
  \BibitemOpen
  \bibfield  {author} {\bibinfo {author} {\bibfnamefont {L.}~\bibnamefont
  {Fern{\'a}ndez}}, \bibinfo {author} {\bibfnamefont {M.}~\bibnamefont
  {Blanco-Rey}}, \bibinfo {author} {\bibfnamefont {R.}~\bibnamefont
  {Castrillo-Bodero}}, \bibinfo {author} {\bibfnamefont {M.}~\bibnamefont
  {Ilyn}}, \bibinfo {author} {\bibfnamefont {K.}~\bibnamefont {Ali}}, \bibinfo
  {author} {\bibfnamefont {E.}~\bibnamefont {Turco}}, \bibinfo {author}
  {\bibfnamefont {M.}~\bibnamefont {Corso}}, \bibinfo {author} {\bibfnamefont
  {M.}~\bibnamefont {Ormaza}}, \bibinfo {author} {\bibfnamefont
  {P.}~\bibnamefont {Gargiani}}, \bibinfo {author} {\bibfnamefont {M.~A.}\
  \bibnamefont {Valbuena}}, \emph {et~al.},\ }\bibfield  {title} {\bibinfo
  {title} {\textit{Influence of 4f filling on electronic and magnetic
  properties of rare earth-Au surface compounds}},\ }\href
  {https://doi.org/10.1039/D0NR04964F} {\bibfield  {journal} {\bibinfo
  {journal} {Nanoscale}\ }\textbf {\bibinfo {volume} {12}},\ \bibinfo {pages}
  {22258} (\bibinfo {year} {2020})}\BibitemShut {NoStop}%
\bibitem [{\citenamefont {Fornari}\ \emph {et~al.}(2020)\citenamefont
  {Fornari}, \citenamefont {Bentmann}, \citenamefont {Morelh{\~a}o},
  \citenamefont {Peixoto}, \citenamefont {Rappl}, \citenamefont {Tcakaev},
  \citenamefont {Zabolotnyy}, \citenamefont {Kamp}, \citenamefont {Lee},
  \citenamefont {Min} \emph {et~al.}}]{fornari2020incorporation}%
  \BibitemOpen
  \bibfield  {author} {\bibinfo {author} {\bibfnamefont {C.~I.}\ \bibnamefont
  {Fornari}}, \bibinfo {author} {\bibfnamefont {H.}~\bibnamefont {Bentmann}},
  \bibinfo {author} {\bibfnamefont {S.~L.}\ \bibnamefont {Morelh{\~a}o}},
  \bibinfo {author} {\bibfnamefont {T.~R.}\ \bibnamefont {Peixoto}}, \bibinfo
  {author} {\bibfnamefont {P.~H.}\ \bibnamefont {Rappl}}, \bibinfo {author}
  {\bibfnamefont {A.-V.}\ \bibnamefont {Tcakaev}}, \bibinfo {author}
  {\bibfnamefont {V.}~\bibnamefont {Zabolotnyy}}, \bibinfo {author}
  {\bibfnamefont {M.}~\bibnamefont {Kamp}}, \bibinfo {author} {\bibfnamefont
  {T.-L.}\ \bibnamefont {Lee}}, \bibinfo {author} {\bibfnamefont {C.-H.}\
  \bibnamefont {Min}}, \emph {et~al.},\ }\bibfield  {title} {\bibinfo {title}
  {\textit{Incorporation of europium in Bi$_{2}$Te$_{3}$ topological insulator
  epitaxial films}},\ }\href {https://doi.org/10.1021/acs.jpcc.0c05077}
  {\bibfield  {journal} {\bibinfo  {journal} {The Journal of Physical Chemistry
  C}\ }\textbf {\bibinfo {volume} {124}},\ \bibinfo {pages} {16048} (\bibinfo
  {year} {2020})}\BibitemShut {NoStop}%
\bibitem [{\citenamefont {Tcakaev}\ \emph {et~al.}(2020)\citenamefont
  {Tcakaev}, \citenamefont {Zabolotnyy}, \citenamefont {Fornari}, \citenamefont
  {R{\"u}{\ss}mann}, \citenamefont {Peixoto}, \citenamefont {Stier},
  \citenamefont {Dettbarn}, \citenamefont {Kagerer}, \citenamefont {Weschke},
  \citenamefont {Schierle} \emph {et~al.}}]{tcakaev2020incipient}%
  \BibitemOpen
  \bibfield  {author} {\bibinfo {author} {\bibfnamefont {A.}~\bibnamefont
  {Tcakaev}}, \bibinfo {author} {\bibfnamefont {V.~B.}\ \bibnamefont
  {Zabolotnyy}}, \bibinfo {author} {\bibfnamefont {C.~I.}\ \bibnamefont
  {Fornari}}, \bibinfo {author} {\bibfnamefont {P.}~\bibnamefont
  {R{\"u}{\ss}mann}}, \bibinfo {author} {\bibfnamefont {T.~R.}\ \bibnamefont
  {Peixoto}}, \bibinfo {author} {\bibfnamefont {F.}~\bibnamefont {Stier}},
  \bibinfo {author} {\bibfnamefont {M.}~\bibnamefont {Dettbarn}}, \bibinfo
  {author} {\bibfnamefont {P.}~\bibnamefont {Kagerer}}, \bibinfo {author}
  {\bibfnamefont {E.}~\bibnamefont {Weschke}}, \bibinfo {author} {\bibfnamefont
  {E.}~\bibnamefont {Schierle}}, \emph {et~al.},\ }\bibfield  {title} {\bibinfo
  {title} {\textit{Incipient antiferromagnetism in the Eu-doped topological
  insulator Bi$_{2} 2$Te$_{3}$}},\ }\href
  {https://doi.org/10.1103/PhysRevB.102.184401} {\bibfield  {journal} {\bibinfo
   {journal} {Physical Review B}\ }\textbf {\bibinfo {volume} {102}},\ \bibinfo
  {pages} {184401} (\bibinfo {year} {2020})}\BibitemShut {NoStop}%
\bibitem [{\citenamefont {Kim}\ \emph {et~al.}(2015)\citenamefont {Kim},
  \citenamefont {Lee}, \citenamefont {Takabatake}, \citenamefont {Kim},
  \citenamefont {Kim},\ and\ \citenamefont {Jung}}]{kim2015magnetic}%
  \BibitemOpen
  \bibfield  {author} {\bibinfo {author} {\bibfnamefont {J.}~\bibnamefont
  {Kim}}, \bibinfo {author} {\bibfnamefont {K.}~\bibnamefont {Lee}}, \bibinfo
  {author} {\bibfnamefont {T.}~\bibnamefont {Takabatake}}, \bibinfo {author}
  {\bibfnamefont {H.}~\bibnamefont {Kim}}, \bibinfo {author} {\bibfnamefont
  {M.}~\bibnamefont {Kim}},\ and\ \bibinfo {author} {\bibfnamefont {M.-H.}\
  \bibnamefont {Jung}},\ }\bibfield  {title} {\bibinfo {title}
  {\textit{Magnetic transition to antiferromagnetic phase in gadolinium
  substituted topological insulator Bi$_{2}$Te$_{3}$}},\ }\href
  {https://doi.org/doi.org/10.1038/srep10309} {\bibfield  {journal} {\bibinfo
  {journal} {Scientific Reports}\ }\textbf {\bibinfo {volume} {5}},\ \bibinfo
  {pages} {10309} (\bibinfo {year} {2015})}\BibitemShut {NoStop}%
\bibitem [{\citenamefont {Harrison}\ \emph
  {et~al.}(2015{\natexlab{b}})\citenamefont {Harrison}, \citenamefont
  {Collins-McIntyre}, \citenamefont {Sch{\"o}nherr}, \citenamefont {Vailionis},
  \citenamefont {Srot}, \citenamefont {van Aken}, \citenamefont {Kellock},
  \citenamefont {Pushp}, \citenamefont {Parkin}, \citenamefont {Harris} \emph
  {et~al.}}]{harrison2015massive}%
  \BibitemOpen
  \bibfield  {author} {\bibinfo {author} {\bibfnamefont {S.}~\bibnamefont
  {Harrison}}, \bibinfo {author} {\bibfnamefont {L.~J.}\ \bibnamefont
  {Collins-McIntyre}}, \bibinfo {author} {\bibfnamefont {P.}~\bibnamefont
  {Sch{\"o}nherr}}, \bibinfo {author} {\bibfnamefont {A.}~\bibnamefont
  {Vailionis}}, \bibinfo {author} {\bibfnamefont {V.}~\bibnamefont {Srot}},
  \bibinfo {author} {\bibfnamefont {P.~A.}\ \bibnamefont {van Aken}}, \bibinfo
  {author} {\bibfnamefont {A.}~\bibnamefont {Kellock}}, \bibinfo {author}
  {\bibfnamefont {A.}~\bibnamefont {Pushp}}, \bibinfo {author} {\bibfnamefont
  {S.}~\bibnamefont {Parkin}}, \bibinfo {author} {\bibfnamefont
  {J.}~\bibnamefont {Harris}}, \emph {et~al.},\ }\bibfield  {title} {\bibinfo
  {title} {\textit{Massive Dirac fermion observed in lanthanide-doped
  topological insulator thin films}},\ }\href
  {https://doi.org/10.1038/srep15767} {\bibfield  {journal} {\bibinfo
  {journal} {Scientific reports}\ }\textbf {\bibinfo {volume} {5}},\ \bibinfo
  {pages} {15767} (\bibinfo {year} {2015}{\natexlab{b}})}\BibitemShut {NoStop}%
\bibitem [{\citenamefont {Hesjedal}(2019)}]{hesjedal2019rare}%
  \BibitemOpen
  \bibfield  {author} {\bibinfo {author} {\bibfnamefont {T.}~\bibnamefont
  {Hesjedal}},\ }\bibfield  {title} {\bibinfo {title} {\textit{Rare earth
  doping of topological insulators: A brief review of thin film and
  heterostructure systems}},\ }\href {https://doi.org/10.1002/pssa.201800726}
  {\bibfield  {journal} {\bibinfo  {journal} {Physica Status Solidi A}\
  }\textbf {\bibinfo {volume} {216}},\ \bibinfo {pages} {1800726} (\bibinfo
  {year} {2019})}\BibitemShut {NoStop}%
\bibitem [{\citenamefont {Figueroa}\ \emph {et~al.}(2020)\citenamefont
  {Figueroa}, \citenamefont {Bonell}, \citenamefont {Cuxart}, \citenamefont
  {Valvidares}, \citenamefont {Gargiani}, \citenamefont {van~der Laan},
  \citenamefont {Mugarza},\ and\ \citenamefont
  {Valenzuela}}]{figueroa2020absence}%
  \BibitemOpen
  \bibfield  {author} {\bibinfo {author} {\bibfnamefont {A.~I.}\ \bibnamefont
  {Figueroa}}, \bibinfo {author} {\bibfnamefont {F.}~\bibnamefont {Bonell}},
  \bibinfo {author} {\bibfnamefont {M.}~\bibnamefont {Cuxart}}, \bibinfo
  {author} {\bibfnamefont {M.}~\bibnamefont {Valvidares}}, \bibinfo {author}
  {\bibfnamefont {P.}~\bibnamefont {Gargiani}}, \bibinfo {author}
  {\bibfnamefont {G.}~\bibnamefont {van~der Laan}}, \bibinfo {author}
  {\bibfnamefont {A.}~\bibnamefont {Mugarza}},\ and\ \bibinfo {author}
  {\bibfnamefont {S.~O.}\ \bibnamefont {Valenzuela}},\ }\bibfield  {title}
  {\bibinfo {title} {\textit{Absence of Magnetic Proximity Effect at the
  Interface of Bi$_{2}$Se$_{3}$ and (Bi, Sb)$_{2}$Te$_{3}$ with EuS}},\ }\href
  {https://doi.org/10.1103/PhysRevLett.125.226801} {\bibfield  {journal}
  {\bibinfo  {journal} {Physical Review Letters}\ }\textbf {\bibinfo {volume}
  {125}},\ \bibinfo {pages} {226801} (\bibinfo {year} {2020})}\BibitemShut
  {NoStop}%
\bibitem [{\citenamefont {He}(2020)}]{he2020mnbi2te4}%
  \BibitemOpen
  \bibfield  {author} {\bibinfo {author} {\bibfnamefont {K.}~\bibnamefont
  {He}},\ }\bibfield  {title} {\bibinfo {title}
  {\textit{MnBi$_{2}$Te$_{4}$-family intrinsic magnetic topological
  materials}},\ }\href {https://doi.org/10.1038/s41535-020-00291-5} {\bibfield
  {journal} {\bibinfo  {journal} {npj Quantum Materials}\ }\textbf {\bibinfo
  {volume} {5}},\ \bibinfo {pages} {90} (\bibinfo {year} {2020})}\BibitemShut
  {NoStop}%
\bibitem [{\citenamefont {Imai}\ \emph {et~al.}(2021)\citenamefont {Imai},
  \citenamefont {Yamaguchi}, \citenamefont {Yamakage},\ and\ \citenamefont
  {Kohno}}]{imai2021spintronic}%
  \BibitemOpen
  \bibfield  {author} {\bibinfo {author} {\bibfnamefont {Y.}~\bibnamefont
  {Imai}}, \bibinfo {author} {\bibfnamefont {T.}~\bibnamefont {Yamaguchi}},
  \bibinfo {author} {\bibfnamefont {A.}~\bibnamefont {Yamakage}},\ and\
  \bibinfo {author} {\bibfnamefont {H.}~\bibnamefont {Kohno}},\ }\bibfield
  {title} {\bibinfo {title} {\textit{Spintronic properties of topological
  surface Dirac electrons with hexagonal warping}},\ }\href
  {https://doi.org/10.1103/PhysRevB.103.054402} {\bibfield  {journal} {\bibinfo
   {journal} {Physical Review B}\ }\textbf {\bibinfo {volume} {103}},\ \bibinfo
  {pages} {054402} (\bibinfo {year} {2021})}\BibitemShut {NoStop}%
\bibitem [{\citenamefont {Naselli}\ \emph {et~al.}(2022)\citenamefont
  {Naselli}, \citenamefont {Moghaddam}, \citenamefont {Di~Napoli},
  \citenamefont {Vildosola}, \citenamefont {Fulga}, \citenamefont {van~den
  Brink},\ and\ \citenamefont {Facio}}]{naselli2022magnetic}%
  \BibitemOpen
  \bibfield  {author} {\bibinfo {author} {\bibfnamefont {G.}~\bibnamefont
  {Naselli}}, \bibinfo {author} {\bibfnamefont {A.~G.}\ \bibnamefont
  {Moghaddam}}, \bibinfo {author} {\bibfnamefont {S.}~\bibnamefont
  {Di~Napoli}}, \bibinfo {author} {\bibfnamefont {V.}~\bibnamefont
  {Vildosola}}, \bibinfo {author} {\bibfnamefont {I.~C.}\ \bibnamefont
  {Fulga}}, \bibinfo {author} {\bibfnamefont {J.}~\bibnamefont {van~den
  Brink}},\ and\ \bibinfo {author} {\bibfnamefont {J.~I.}\ \bibnamefont
  {Facio}},\ }\bibfield  {title} {\bibinfo {title} {\textit{Magnetic warping in
  topological insulators}},\ }\href
  {https://doi.org/10.1103/PhysRevResearch.4.033198} {\bibfield  {journal}
  {\bibinfo  {journal} {Physical Review Research}\ }\textbf {\bibinfo {volume}
  {4}},\ \bibinfo {pages} {033198} (\bibinfo {year} {2022})}\BibitemShut
  {NoStop}%
\bibitem [{\citenamefont {Tan}\ \emph {et~al.}(2022)\citenamefont {Tan},
  \citenamefont {Kaplan},\ and\ \citenamefont {Yan}}]{tan2022momentum}%
  \BibitemOpen
  \bibfield  {author} {\bibinfo {author} {\bibfnamefont {H.}~\bibnamefont
  {Tan}}, \bibinfo {author} {\bibfnamefont {D.}~\bibnamefont {Kaplan}},\ and\
  \bibinfo {author} {\bibfnamefont {B.}~\bibnamefont {Yan}},\ }\bibfield
  {title} {\bibinfo {title} {\textit{Momentum-inversion symmetry breaking on
  the Fermi surface of magnetic topological insulators}},\ }\href
  {https://doi.org/10.1103/PhysRevMaterials.6.104204} {\bibfield  {journal}
  {\bibinfo  {journal} {Physical Review Materials}\ }\textbf {\bibinfo {volume}
  {6}},\ \bibinfo {pages} {104204} (\bibinfo {year} {2022})}\BibitemShut
  {NoStop}%
\bibitem [{\citenamefont {Wang}\ \emph {et~al.}(2022)\citenamefont {Wang},
  \citenamefont {Wang}, \citenamefont {Xing},\ and\ \citenamefont
  {Zhang}}]{wang2022three}%
  \BibitemOpen
  \bibfield  {author} {\bibinfo {author} {\bibfnamefont {D.}~\bibnamefont
  {Wang}}, \bibinfo {author} {\bibfnamefont {H.}~\bibnamefont {Wang}}, \bibinfo
  {author} {\bibfnamefont {D.}~\bibnamefont {Xing}},\ and\ \bibinfo {author}
  {\bibfnamefont {H.}~\bibnamefont {Zhang}},\ }\bibfield  {title} {\bibinfo
  {title} {\textit{Three-Dirac-fermion approach to unexpected gapless surface
  states of van der Waals magnetic topological insulators}},\ }\href
  {https://arxiv.org/abs/2205.08204} {\bibfield  {journal} {\bibinfo  {journal}
  {arXiv preprint arXiv:2205.08204}\ } (\bibinfo {year} {2022})}\BibitemShut
  {NoStop}%
\bibitem [{\citenamefont {Wang}\ and\ \citenamefont
  {Johnson}(2011)}]{wang2011ternary}%
  \BibitemOpen
  \bibfield  {author} {\bibinfo {author} {\bibfnamefont {L.-L.}\ \bibnamefont
  {Wang}}\ and\ \bibinfo {author} {\bibfnamefont {D.~D.}\ \bibnamefont
  {Johnson}},\ }\bibfield  {title} {\bibinfo {title} {\textit{Ternary
  tetradymite compounds as topological insulators}},\ }\href
  {https://doi.org/10.1103/PhysRevB.83.241309} {\bibfield  {journal} {\bibinfo
  {journal} {Physical Review B}\ }\textbf {\bibinfo {volume} {83}},\ \bibinfo
  {pages} {241309} (\bibinfo {year} {2011})}\BibitemShut {NoStop}%
\bibitem [{sm()}]{sm}%
  \BibitemOpen
  \href@noop {} {\bibinfo {title} {See {S}upplemental {M}aterials for the
  description of the growth process, the {Er} coverage calculation, further
  {XPS} and {ARPES} measurements and the derivation of the theoretical
  model}},\ \bibinfo {howpublished}
  {\url{URL_will_be_inserted_by_publisher}}\BibitemShut {NoStop}%
\bibitem [{\citenamefont {Garc{\'\i}a}\ \emph {et~al.}(2022)\citenamefont
  {Garc{\'\i}a}, \citenamefont {Martin}, \citenamefont {Ynsa}, \citenamefont
  {Torres-Costa}, \citenamefont {Crespillo}, \citenamefont {Tard{\'\i}o},
  \citenamefont {Olivares}, \citenamefont {Bosia}, \citenamefont
  {Pe{\~n}a-Rodr{\'\i}guez}, \citenamefont {Nicolas} \emph
  {et~al.}}]{garcia2022process}%
  \BibitemOpen
  \bibfield  {author} {\bibinfo {author} {\bibfnamefont {G.}~\bibnamefont
  {Garc{\'\i}a}}, \bibinfo {author} {\bibfnamefont {M.}~\bibnamefont {Martin}},
  \bibinfo {author} {\bibfnamefont {M.}~\bibnamefont {Ynsa}}, \bibinfo {author}
  {\bibfnamefont {V.}~\bibnamefont {Torres-Costa}}, \bibinfo {author}
  {\bibfnamefont {M.}~\bibnamefont {Crespillo}}, \bibinfo {author}
  {\bibfnamefont {M.}~\bibnamefont {Tard{\'\i}o}}, \bibinfo {author}
  {\bibfnamefont {J.}~\bibnamefont {Olivares}}, \bibinfo {author}
  {\bibfnamefont {F.}~\bibnamefont {Bosia}}, \bibinfo {author} {\bibfnamefont
  {O.}~\bibnamefont {Pe{\~n}a-Rodr{\'\i}guez}}, \bibinfo {author}
  {\bibfnamefont {J.}~\bibnamefont {Nicolas}}, \emph {et~al.},\ }\bibfield
  {title} {\bibinfo {title} {\textit{Process design for the manufacturing of
  soft X-ray gratings in single-crystal diamond by high-energy heavy-ion
  irradiation}},\ }\href {https://doi.org/10.1140/epjp/s13360-022-03358-3}
  {\bibfield  {journal} {\bibinfo  {journal} {The European Physical Journal
  Plus}\ }\textbf {\bibinfo {volume} {137}},\ \bibinfo {pages} {1157} (\bibinfo
  {year} {2022})}\BibitemShut {NoStop}%
\bibitem [{\citenamefont {Crisol}\ \emph {et~al.}(2021)\citenamefont {Crisol},
  \citenamefont {Bisti}, \citenamefont {Colldelram}, \citenamefont {Llonch},
  \citenamefont {Molas}, \citenamefont {Monge}, \citenamefont {Nicolás},
  \citenamefont {Nikitina}, \citenamefont {Quispe}, \citenamefont {Ribó},\
  and\ \citenamefont {Tallarida}}]{crisol2021alba}%
  \BibitemOpen
  \bibfield  {author} {\bibinfo {author} {\bibfnamefont {A.}~\bibnamefont
  {Crisol}}, \bibinfo {author} {\bibfnamefont {F.}~\bibnamefont {Bisti}},
  \bibinfo {author} {\bibfnamefont {C.}~\bibnamefont {Colldelram}}, \bibinfo
  {author} {\bibfnamefont {M.}~\bibnamefont {Llonch}}, \bibinfo {author}
  {\bibfnamefont {B.}~\bibnamefont {Molas}}, \bibinfo {author} {\bibfnamefont
  {R.}~\bibnamefont {Monge}}, \bibinfo {author} {\bibfnamefont
  {J.}~\bibnamefont {Nicolás}}, \bibinfo {author} {\bibfnamefont
  {L.}~\bibnamefont {Nikitina}}, \bibinfo {author} {\bibfnamefont
  {M.}~\bibnamefont {Quispe}}, \bibinfo {author} {\bibfnamefont
  {L.}~\bibnamefont {Ribó}},\ and\ \bibinfo {author} {\bibfnamefont
  {M.}~\bibnamefont {Tallarida}},\ }\bibfield  {title} {\bibinfo {title}
  {\textit{ALBA BL20 New Monochromator Design}},\ }in\ \href
  {https://doi.org/10.18429/JACoW-MEDSI2020-MOOB02} {\emph {\bibinfo
  {booktitle} {Proc. MEDSI'20}}},\ \bibinfo {series and number} {\bibinfo
  {series} {Mechanical Engineering Design of Synchrotron Radiation Equipment
  and Instrumentation}\ No.~\bibinfo {number} {11}}\ (\bibinfo  {publisher}
  {JACoW Publishing, Geneva, Switzerland},\ \bibinfo {year} {2021})\ pp.\
  \bibinfo {pages} {14--16}\BibitemShut {NoStop}%
\bibitem [{\citenamefont {Banister}\ \emph {et~al.}(1954)\citenamefont
  {Banister}, \citenamefont {Legvold},\ and\ \citenamefont
  {Spedding}}]{banister1954structure}%
  \BibitemOpen
  \bibfield  {author} {\bibinfo {author} {\bibfnamefont {J.}~\bibnamefont
  {Banister}}, \bibinfo {author} {\bibfnamefont {S.}~\bibnamefont {Legvold}},\
  and\ \bibinfo {author} {\bibfnamefont {F.}~\bibnamefont {Spedding}},\
  }\bibfield  {title} {\bibinfo {title} {\textit{Structure of Gd, Dy, and Er at
  low temperatures}},\ }\href {https://doi.org/10.1103/PhysRev.94.1140}
  {\bibfield  {journal} {\bibinfo  {journal} {Physical Review}\ }\textbf
  {\bibinfo {volume} {94}},\ \bibinfo {pages} {1140} (\bibinfo {year}
  {1954})}\BibitemShut {NoStop}%
\bibitem [{\citenamefont {Maa{\ss}}\ \emph {et~al.}(2014)\citenamefont
  {Maa{\ss}}, \citenamefont {Schreyeck}, \citenamefont {Schatz}, \citenamefont
  {Fiedler}, \citenamefont {Seibel}, \citenamefont {Lutz}, \citenamefont
  {Karczewski}, \citenamefont {Bentmann}, \citenamefont {Gould}, \citenamefont
  {Brunner} \emph {et~al.}}]{maass2014electronic}%
  \BibitemOpen
  \bibfield  {author} {\bibinfo {author} {\bibfnamefont {H.}~\bibnamefont
  {Maa{\ss}}}, \bibinfo {author} {\bibfnamefont {S.}~\bibnamefont {Schreyeck}},
  \bibinfo {author} {\bibfnamefont {S.}~\bibnamefont {Schatz}}, \bibinfo
  {author} {\bibfnamefont {S.}~\bibnamefont {Fiedler}}, \bibinfo {author}
  {\bibfnamefont {C.}~\bibnamefont {Seibel}}, \bibinfo {author} {\bibfnamefont
  {P.}~\bibnamefont {Lutz}}, \bibinfo {author} {\bibfnamefont {G.}~\bibnamefont
  {Karczewski}}, \bibinfo {author} {\bibfnamefont {H.}~\bibnamefont
  {Bentmann}}, \bibinfo {author} {\bibfnamefont {C.}~\bibnamefont {Gould}},
  \bibinfo {author} {\bibfnamefont {K.}~\bibnamefont {Brunner}}, \emph
  {et~al.},\ }\bibfield  {title} {\bibinfo {title} {\textit{Electronic
  structure and morphology of epitaxial Bi$_{2}$Te$_{2}$Se topological
  insulator films}},\ }\href {https://doi.org/10.1063/1.4902010} {\bibfield
  {journal} {\bibinfo  {journal} {Journal of Applied Physics}\ }\textbf
  {\bibinfo {volume} {116}},\ \bibinfo {pages} {193708} (\bibinfo {year}
  {2014})}\BibitemShut {NoStop}%
\bibitem [{\citenamefont {Walsh}\ \emph {et~al.}(2017)\citenamefont {Walsh},
  \citenamefont {Smyth}, \citenamefont {Barton}, \citenamefont {Wang},
  \citenamefont {Che}, \citenamefont {Yue}, \citenamefont {Kim}, \citenamefont
  {Kim}, \citenamefont {Wallace},\ and\ \citenamefont
  {Hinkle}}]{walsh2017interface}%
  \BibitemOpen
  \bibfield  {author} {\bibinfo {author} {\bibfnamefont {L.~A.}\ \bibnamefont
  {Walsh}}, \bibinfo {author} {\bibfnamefont {C.~M.}\ \bibnamefont {Smyth}},
  \bibinfo {author} {\bibfnamefont {A.~T.}\ \bibnamefont {Barton}}, \bibinfo
  {author} {\bibfnamefont {Q.}~\bibnamefont {Wang}}, \bibinfo {author}
  {\bibfnamefont {Z.}~\bibnamefont {Che}}, \bibinfo {author} {\bibfnamefont
  {R.}~\bibnamefont {Yue}}, \bibinfo {author} {\bibfnamefont {J.}~\bibnamefont
  {Kim}}, \bibinfo {author} {\bibfnamefont {M.~J.}\ \bibnamefont {Kim}},
  \bibinfo {author} {\bibfnamefont {R.~M.}\ \bibnamefont {Wallace}},\ and\
  \bibinfo {author} {\bibfnamefont {C.~L.}\ \bibnamefont {Hinkle}},\ }\bibfield
   {title} {\bibinfo {title} {\textit{Interface chemistry of contact metals and
  ferromagnets on the topological insulator Bi$_{2}$Se$_{3}$}},\ }\href
  {https://doi.org/10.1021/acs.jpcc.7b08480} {\bibfield  {journal} {\bibinfo
  {journal} {The Journal of Physical Chemistry C}\ }\textbf {\bibinfo {volume}
  {121}},\ \bibinfo {pages} {23551} (\bibinfo {year} {2017})}\BibitemShut
  {NoStop}%
\bibitem [{\citenamefont {Miyamoto}\ \emph {et~al.}(2012)\citenamefont
  {Miyamoto}, \citenamefont {Kimura}, \citenamefont {Okuda}, \citenamefont
  {Miyahara}, \citenamefont {Kuroda}, \citenamefont {Namatame}, \citenamefont
  {Taniguchi}, \citenamefont {Eremeev}, \citenamefont {Menshchikova},
  \citenamefont {Chulkov} \emph {et~al.}}]{miyamoto2012topological}%
  \BibitemOpen
  \bibfield  {author} {\bibinfo {author} {\bibfnamefont {K.}~\bibnamefont
  {Miyamoto}}, \bibinfo {author} {\bibfnamefont {A.}~\bibnamefont {Kimura}},
  \bibinfo {author} {\bibfnamefont {T.}~\bibnamefont {Okuda}}, \bibinfo
  {author} {\bibfnamefont {H.}~\bibnamefont {Miyahara}}, \bibinfo {author}
  {\bibfnamefont {K.}~\bibnamefont {Kuroda}}, \bibinfo {author} {\bibfnamefont
  {H.}~\bibnamefont {Namatame}}, \bibinfo {author} {\bibfnamefont
  {M.}~\bibnamefont {Taniguchi}}, \bibinfo {author} {\bibfnamefont
  {S.}~\bibnamefont {Eremeev}}, \bibinfo {author} {\bibfnamefont {T.~V.}\
  \bibnamefont {Menshchikova}}, \bibinfo {author} {\bibfnamefont {E.~V.}\
  \bibnamefont {Chulkov}}, \emph {et~al.},\ }\bibfield  {title} {\bibinfo
  {title} {\textit{Topological surface states with persistent high spin
  polarization across the Dirac point in Bi$_{2}$Te$_{2}$Se and
  Bi$_{2}$Se$_{2}$Te}},\ }\href
  {https://doi.org/10.1103/PhysRevLett.109.166802} {\bibfield  {journal}
  {\bibinfo  {journal} {Physical Review Letters}\ }\textbf {\bibinfo {volume}
  {109}},\ \bibinfo {pages} {166802} (\bibinfo {year} {2012})}\BibitemShut
  {NoStop}%
\bibitem [{\citenamefont {Bao}\ \emph {et~al.}(2012)\citenamefont {Bao},
  \citenamefont {He}, \citenamefont {Meyer}, \citenamefont {Kou}, \citenamefont
  {Zhang}, \citenamefont {Chen}, \citenamefont {Fedorov}, \citenamefont {Zou},
  \citenamefont {Riedemann}, \citenamefont {Lograsso} \emph
  {et~al.}}]{bao2012weakARPES}%
  \BibitemOpen
  \bibfield  {author} {\bibinfo {author} {\bibfnamefont {L.}~\bibnamefont
  {Bao}}, \bibinfo {author} {\bibfnamefont {L.}~\bibnamefont {He}}, \bibinfo
  {author} {\bibfnamefont {N.}~\bibnamefont {Meyer}}, \bibinfo {author}
  {\bibfnamefont {X.}~\bibnamefont {Kou}}, \bibinfo {author} {\bibfnamefont
  {P.}~\bibnamefont {Zhang}}, \bibinfo {author} {\bibfnamefont {Z.-g.}\
  \bibnamefont {Chen}}, \bibinfo {author} {\bibfnamefont {A.~V.}\ \bibnamefont
  {Fedorov}}, \bibinfo {author} {\bibfnamefont {J.}~\bibnamefont {Zou}},
  \bibinfo {author} {\bibfnamefont {T.~M.}\ \bibnamefont {Riedemann}}, \bibinfo
  {author} {\bibfnamefont {T.~A.}\ \bibnamefont {Lograsso}}, \emph {et~al.},\
  }\bibfield  {title} {\bibinfo {title} {\textit{Weak anti-localization and
  quantum oscillations of surface states in topological insulator
  Bi$_{2}$Se$_{2}$Te}},\ }\href {https://doi.org/10.1038/srep00726} {\bibfield
  {journal} {\bibinfo  {journal} {Scientific Reports}\ }\textbf {\bibinfo
  {volume} {2}},\ \bibinfo {pages} {726} (\bibinfo {year} {2012})}\BibitemShut
  {NoStop}%
\bibitem [{\citenamefont {Fu}(2009)}]{fu2009hexagonal}%
  \BibitemOpen
  \bibfield  {author} {\bibinfo {author} {\bibfnamefont {L.}~\bibnamefont
  {Fu}},\ }\bibfield  {title} {\bibinfo {title} {\textit{Hexagonal warping
  effects in the surface states of the topological insulator
  Bi$_{2}$Te$_{3}$}},\ }\href {https://doi.org/10.1103/PhysRevLett.103.266801}
  {\bibfield  {journal} {\bibinfo  {journal} {Physical Review Letters}\
  }\textbf {\bibinfo {volume} {103}},\ \bibinfo {pages} {266801} (\bibinfo
  {year} {2009})}\BibitemShut {NoStop}%
\bibitem [{\citenamefont {Chen}\ \emph {et~al.}(2009)\citenamefont {Chen},
  \citenamefont {Analytis}, \citenamefont {Chu}, \citenamefont {Liu},
  \citenamefont {Mo}, \citenamefont {Qi}, \citenamefont {Zhang}, \citenamefont
  {Lu}, \citenamefont {Dai}, \citenamefont {Fang} \emph
  {et~al.}}]{chen2009experimental}%
  \BibitemOpen
  \bibfield  {author} {\bibinfo {author} {\bibfnamefont {Y.}~\bibnamefont
  {Chen}}, \bibinfo {author} {\bibfnamefont {J.~G.}\ \bibnamefont {Analytis}},
  \bibinfo {author} {\bibfnamefont {J.-H.}\ \bibnamefont {Chu}}, \bibinfo
  {author} {\bibfnamefont {Z.}~\bibnamefont {Liu}}, \bibinfo {author}
  {\bibfnamefont {S.-K.}\ \bibnamefont {Mo}}, \bibinfo {author} {\bibfnamefont
  {X.-L.}\ \bibnamefont {Qi}}, \bibinfo {author} {\bibfnamefont
  {H.}~\bibnamefont {Zhang}}, \bibinfo {author} {\bibfnamefont
  {D.}~\bibnamefont {Lu}}, \bibinfo {author} {\bibfnamefont {X.}~\bibnamefont
  {Dai}}, \bibinfo {author} {\bibfnamefont {Z.}~\bibnamefont {Fang}}, \emph
  {et~al.},\ }\bibfield  {title} {\bibinfo {title} {\textit{Experimental
  realization of a three-dimensional topological insulator,
  Bi$_{2}$Te$_{3}$}},\ }\href {https://doi.org/10.1126/science.117303}
  {\bibfield  {journal} {\bibinfo  {journal} {Science}\ }\textbf {\bibinfo
  {volume} {325}},\ \bibinfo {pages} {178} (\bibinfo {year}
  {2009})}\BibitemShut {NoStop}%
\bibitem [{\citenamefont {Arakane}\ \emph {et~al.}(2012)\citenamefont
  {Arakane}, \citenamefont {Sato}, \citenamefont {Souma}, \citenamefont
  {Kosaka}, \citenamefont {Nakayama}, \citenamefont {Komatsu}, \citenamefont
  {Takahashi}, \citenamefont {Ren}, \citenamefont {Segawa},\ and\ \citenamefont
  {Ando}}]{arakane2012tunable}%
  \BibitemOpen
  \bibfield  {author} {\bibinfo {author} {\bibfnamefont {T.}~\bibnamefont
  {Arakane}}, \bibinfo {author} {\bibfnamefont {T.}~\bibnamefont {Sato}},
  \bibinfo {author} {\bibfnamefont {S.}~\bibnamefont {Souma}}, \bibinfo
  {author} {\bibfnamefont {K.}~\bibnamefont {Kosaka}}, \bibinfo {author}
  {\bibfnamefont {K.}~\bibnamefont {Nakayama}}, \bibinfo {author}
  {\bibfnamefont {M.}~\bibnamefont {Komatsu}}, \bibinfo {author} {\bibfnamefont
  {T.}~\bibnamefont {Takahashi}}, \bibinfo {author} {\bibfnamefont
  {Z.}~\bibnamefont {Ren}}, \bibinfo {author} {\bibfnamefont {K.}~\bibnamefont
  {Segawa}},\ and\ \bibinfo {author} {\bibfnamefont {Y.}~\bibnamefont {Ando}},\
  }\bibfield  {title} {\bibinfo {title} {\textit{Tunable Dirac cone in the
  topological insulator Bi$_{2-x}$Sb$_{x}$Te$_{3-y}$Se$_{y}$}},\ }\href
  {https://doi.org/10.1038/ncomms1639} {\bibfield  {journal} {\bibinfo
  {journal} {Nature Communications}\ }\textbf {\bibinfo {volume} {3}},\
  \bibinfo {pages} {636} (\bibinfo {year} {2012})}\BibitemShut {NoStop}%
\bibitem [{\citenamefont {Xia}\ \emph {et~al.}(2009)\citenamefont {Xia},
  \citenamefont {Qian}, \citenamefont {Hsieh}, \citenamefont {Wray},
  \citenamefont {Pal}, \citenamefont {Lin}, \citenamefont {Bansil},
  \citenamefont {Grauer}, \citenamefont {Hor}, \citenamefont {Cava} \emph
  {et~al.}}]{xia2009observation}%
  \BibitemOpen
  \bibfield  {author} {\bibinfo {author} {\bibfnamefont {Y.}~\bibnamefont
  {Xia}}, \bibinfo {author} {\bibfnamefont {D.}~\bibnamefont {Qian}}, \bibinfo
  {author} {\bibfnamefont {D.}~\bibnamefont {Hsieh}}, \bibinfo {author}
  {\bibfnamefont {L.}~\bibnamefont {Wray}}, \bibinfo {author} {\bibfnamefont
  {A.}~\bibnamefont {Pal}}, \bibinfo {author} {\bibfnamefont {H.}~\bibnamefont
  {Lin}}, \bibinfo {author} {\bibfnamefont {A.}~\bibnamefont {Bansil}},
  \bibinfo {author} {\bibfnamefont {D.}~\bibnamefont {Grauer}}, \bibinfo
  {author} {\bibfnamefont {Y.~S.}\ \bibnamefont {Hor}}, \bibinfo {author}
  {\bibfnamefont {R.~J.}\ \bibnamefont {Cava}}, \emph {et~al.},\ }\bibfield
  {title} {\bibinfo {title} {\textit{Observation of a large-gap
  topological-insulator class with a single Dirac cone on the surface}},\
  }\href {https://doi.org/10.1038/nphys1274} {\bibfield  {journal} {\bibinfo
  {journal} {Nature Physics}\ }\textbf {\bibinfo {volume} {5}},\ \bibinfo
  {pages} {398} (\bibinfo {year} {2009})}\BibitemShut {NoStop}%
\bibitem [{\citenamefont {Qi}\ \emph {et~al.}(2006)\citenamefont {Qi},
  \citenamefont {Wu},\ and\ \citenamefont {Zhang}}]{qi2006topological}%
  \BibitemOpen
  \bibfield  {author} {\bibinfo {author} {\bibfnamefont {X.-L.}\ \bibnamefont
  {Qi}}, \bibinfo {author} {\bibfnamefont {Y.-S.}\ \bibnamefont {Wu}},\ and\
  \bibinfo {author} {\bibfnamefont {S.-C.}\ \bibnamefont {Zhang}},\ }\bibfield
  {title} {\bibinfo {title} {\textit{Topological quantization of the spin Hall
  effect in two-dimensional paramagnetic semiconductors}},\ }\href
  {https://doi.org/10.1103/PhysRevB.74.085308} {\bibfield  {journal} {\bibinfo
  {journal} {Physical Review B}\ }\textbf {\bibinfo {volume} {74}},\ \bibinfo
  {pages} {085308} (\bibinfo {year} {2006})}\BibitemShut {NoStop}%
\bibitem [{\citenamefont {Nomura}\ and\ \citenamefont {Nagaosa}(2010)}]{NN10}%
  \BibitemOpen
  \bibfield  {author} {\bibinfo {author} {\bibfnamefont {K.}~\bibnamefont
  {Nomura}}\ and\ \bibinfo {author} {\bibfnamefont {N.}~\bibnamefont
  {Nagaosa}},\ }\bibfield  {title} {\bibinfo {title} {\textit{Electric charging
  of magnetic textures on the surface of a topological insulator}},\ }\href
  {https://doi.org/10.1103/PhysRevB.82.161401} {\bibfield  {journal} {\bibinfo
  {journal} {Physical Review B}\ }\textbf {\bibinfo {volume} {82}},\ \bibinfo
  {pages} {161401} (\bibinfo {year} {2010})}\BibitemShut {NoStop}%
\bibitem [{\citenamefont {Yokoyama}\ \emph {et~al.}(2010)\citenamefont
  {Yokoyama}, \citenamefont {Zang},\ and\ \citenamefont
  {Nagaosa}}]{yokoyama2010theoretical}%
  \BibitemOpen
  \bibfield  {author} {\bibinfo {author} {\bibfnamefont {T.}~\bibnamefont
  {Yokoyama}}, \bibinfo {author} {\bibfnamefont {J.}~\bibnamefont {Zang}},\
  and\ \bibinfo {author} {\bibfnamefont {N.}~\bibnamefont {Nagaosa}},\
  }\bibfield  {title} {\bibinfo {title} {\textit{Theoretical study of the
  dynamics of magnetization on the topological surface}},\ }\href
  {https://doi.org/10.1103/PhysRevB.81.241410} {\bibfield  {journal} {\bibinfo
  {journal} {Physical Review B}\ }\textbf {\bibinfo {volume} {81}},\ \bibinfo
  {pages} {241410} (\bibinfo {year} {2010})}\BibitemShut {NoStop}%
\bibitem [{\citenamefont {Yokoyama}(2011)}]{Y11}%
  \BibitemOpen
  \bibfield  {author} {\bibinfo {author} {\bibfnamefont {T.}~\bibnamefont
  {Yokoyama}},\ }\bibfield  {title} {\bibinfo {title} {\textit{Current-induced
  magnetization reversal on the surface of a topological insulator}},\ }\href
  {https://doi.org/10.1103/PhysRevB.84.113407} {\bibfield  {journal} {\bibinfo
  {journal} {Phys. Rev. B}\ }\textbf {\bibinfo {volume} {84}},\ \bibinfo
  {pages} {113407} (\bibinfo {year} {2011})}\BibitemShut {NoStop}%
\bibitem [{\citenamefont {Ferreiros}\ \emph {et~al.}(2015)\citenamefont
  {Ferreiros}, \citenamefont {Buijnsters},\ and\ \citenamefont
  {Katsnelson}}]{ferreiros2015dirac}%
  \BibitemOpen
  \bibfield  {author} {\bibinfo {author} {\bibfnamefont {Y.}~\bibnamefont
  {Ferreiros}}, \bibinfo {author} {\bibfnamefont {F.}~\bibnamefont
  {Buijnsters}},\ and\ \bibinfo {author} {\bibfnamefont {M.}~\bibnamefont
  {Katsnelson}},\ }\bibfield  {title} {\bibinfo {title} {\textit{Dirac
  electrons and domain walls: A realization in junctions of ferromagnets and
  topological insulators}},\ }\href
  {https://doi.org/10.1103/PhysRevB.92.085416} {\bibfield  {journal} {\bibinfo
  {journal} {Physical Review B}\ }\textbf {\bibinfo {volume} {92}},\ \bibinfo
  {pages} {085416} (\bibinfo {year} {2015})}\BibitemShut {NoStop}%
\bibitem [{\citenamefont {Liu}\ \emph {et~al.}(2009)\citenamefont {Liu},
  \citenamefont {Liu}, \citenamefont {Xu}, \citenamefont {Qi},\ and\
  \citenamefont {Zhang}}]{liu2009magnetic}%
  \BibitemOpen
  \bibfield  {author} {\bibinfo {author} {\bibfnamefont {Q.}~\bibnamefont
  {Liu}}, \bibinfo {author} {\bibfnamefont {C.-X.}\ \bibnamefont {Liu}},
  \bibinfo {author} {\bibfnamefont {C.}~\bibnamefont {Xu}}, \bibinfo {author}
  {\bibfnamefont {X.-L.}\ \bibnamefont {Qi}},\ and\ \bibinfo {author}
  {\bibfnamefont {S.-C.}\ \bibnamefont {Zhang}},\ }\bibfield  {title} {\bibinfo
  {title} {\textit{Magnetic impurities on the surface of a topological
  insulator}},\ }\href {https://doi.org/10.1103/PhysRevLett.102.156603}
  {\bibfield  {journal} {\bibinfo  {journal} {Physical Review Letters}\
  }\textbf {\bibinfo {volume} {102}},\ \bibinfo {pages} {156603} (\bibinfo
  {year} {2009})}\BibitemShut {NoStop}%
\bibitem [{\citenamefont {Ferreiros}\ and\ \citenamefont
  {Cortijo}(2014)}]{FC14}%
  \BibitemOpen
  \bibfield  {author} {\bibinfo {author} {\bibfnamefont {Y.}~\bibnamefont
  {Ferreiros}}\ and\ \bibinfo {author} {\bibfnamefont {A.}~\bibnamefont
  {Cortijo}},\ }\bibfield  {title} {\bibinfo {title} {\textit{Domain wall
  motion in junctions of thin-film magnets and topological insulators}},\
  }\href {https://doi.org/10.1103/PhysRevB.89.024413} {\bibfield  {journal}
  {\bibinfo  {journal} {Physical Review B}\ }\textbf {\bibinfo {volume} {89}},\
  \bibinfo {pages} {024413} (\bibinfo {year} {2014})}\BibitemShut {NoStop}%
\end{thebibliography}%

\end{document}